\documentclass[11pt]{article}

\usepackage{amsmath}
\usepackage{amssymb}
\usepackage{amsthm}
\usepackage{appendix}
\usepackage{xcolor}
\newcommand{\W}{\mathsf{W}}
\newcommand{\FPT}{\mathsf{FPT}}
\usepackage{fullpage}
\usepackage{microtype}
\usepackage{enumitem}
\usepackage{thm-restate}
\usepackage{epigraph}
\usepackage{hyperref}
\hypersetup{
  pdfauthor={},
  colorlinks=true,
  linkcolor={red!75!black},
  citecolor={blue!75!black},
  urlcolor={red!75!black},
}
\usepackage[nameinlink]{cleveref}
\crefname{equation}{Equation}{Equations}
\Crefname{equation}{Equation}{Equations}
\usepackage[
  backend=biber,
  style=alphabetic,
  natbib=false,
  maxnames=99,
  maxalphanames=4,
  minalphanames=3,
  useprefix=true,
  backref=true
]{biblatex}

\theoremstyle{plain}
\newtheorem{theorem}{Theorem}[section]
\newtheorem{lemma}[theorem]{Lemma}

\newtheorem{fact}[theorem]{Fact}
\newtheorem{claim}{Claim}

\theoremstyle{definition}
\newtheorem{definition}[theorem]{Definition}
\theoremstyle{remark}

\newcommand{\F}{\mathbb F}
\newcommand{\Z}{\mathbb Z}

\renewcommand{\vec}[1]{\ensuremath{\boldsymbol{#1}}}
\newcommand{\problem}[1]{\ensuremath{\mathrm{#1}}}

\newcommand{\lat}{\ensuremath{\mathcal{L}}}

\newcommand{\GapSVP}{\problem{GapSVP}}
\newcommand{\GapCVP}{\problem{GapCVP}}
\providecommand{\GapNCP}{\problem{GapNCP}}
\providecommand{\GapMDP}{\problem{GapMDP}}
\providecommand{\GapMLD}{\problem{GapMLD}}

\providecommand{\C}{\mathcal{C}}

\usepackage[draft,multiuser,inline,nomargin]{fixme}
\usepackage[left=1in, right=1in, top=1in]{geometry}
\title{Deterministic Parameterized Inapproximability of \\the Shortest Vector and  Minimum Distance Problems}
\author{Shuichi Hirahara\thanks{National Institute of Informatics, Japan. \href{mailto:s_hirahara@nii.ac.jp}{\texttt{s\_hirahara@nii.ac.jp}}}
\and Bingkai Lin\thanks{Nanjing University, China. \href{mailto:lin@nju.edu.cn}{\texttt{lin@nju.edu.cn}}}
\and Yuwei Liu\thanks{The Graduate University for Advanced Studies (SOKENDAI), Japan. \href{mailto:yuwei_liu@nii.ac.jp}{\texttt{yuwei\_liu@nii.ac.jp}}}
\and Kazuki Ogitsuka\thanks{The Graduate University for Advanced Studies (SOKENDAI), Japan. \href{mailto:ogitsuka@nii.ac.jp}{\texttt{ogitsuka@nii.ac.jp}}}
}
\date{}
\begin{document}
\maketitle

\begin{abstract}
The Shortest Vector Problem (SVP) and the Minimum Distance Problem (MDP) are central computational problems on lattices and linear codes, respectively.
For every computable real $p\ge1$ and any constant factor, we prove that the parameterized SVP in the $\ell_p$ norm, with the distance threshold as the parameter, is $\W[1]$-hard to approximate within factor $\gamma$ under deterministic reductions.
The same method shows that, for every finite field, the parameterized MDP is $\W[1]$-hard to approximate within any constant factor under deterministic reductions.
\end{abstract}

\section{Introduction} \label{sec:intro}

The Shortest Vector Problem (SVP) asks for the minimum $\ell_p$ norm of a nonzero vector in a given integer lattice.
It is the central computational problem on lattices, with applications to algorithms and cryptography~\cite{survey/AgrellEVZ02,survey/Peikert16}.
For finite $\ell_p$ norms, the hardness of SVP was long known only under randomized reductions, and removing this randomness has been a major challenge.

In the classical setting of NP-hardness, this challenge has recently been met.
Van Emde Boas~\cite{vanEmdeBoas} proved that SVP in the $\ell_\infty$ norm is NP-hard and conjectured the same for the Euclidean norm.
Ajtai~\cite{conf/stoc/Ajtai98} proved this conjecture under randomized reductions, and subsequent work proved the hardness of approximating SVP in every finite $\ell_p$ norm, again under randomized reductions~\cite{journals/sicomp/Micciancio01,journals/JACM/Kho05,journals/toc/HR12,journals/toc/Micciancio12}.
The first deterministic hardness results for SVP in finite $\ell_p$ norms appeared only recently.
For every $p>2$, Hair and Sahai~\cite{conf/stoc/HairS26} proved that approximating SVP is NP-hard under deterministic polynomial-time reductions, later even within polynomial factors~\cite{HairSahai2026Polynomial}, and Hecht and Safra~\cite{conf/stoc/HechtS26} proved hardness under deterministic subexponential-time reductions.
Hair and Sahai~\cite{hair2026finite} then extended hardness under subexponential-time reductions to every finite $\ell_p$ norm.
Wan~\cite{wan2026} resolved the conjecture of van Emde Boas: for every $p\ge1$, approximating SVP in the $\ell_p$ norm within any factor below $2^{1/p}$ is NP-hard under deterministic polynomial-time reductions.
Wan~\cite{wan2026anyconstant} later extended this result to every constant factor in every finite $\ell_p$ norm.

The situation is, however, different in parameterized complexity.
A problem with a parameter $k$ is \emph{fixed-parameter tractable} (FPT) if it can be solved in time $f(k)\cdot n^{O(1)}$ for some computable function $f$, where $n$ is the input length.
The standard evidence that a problem is not FPT is its $\W[1]$-hardness under FPT reductions, which plays the role of NP-hardness~\cite{book/DowneyF99,book/CyganFKLMPPS15}.
For SVP, the natural parameter is the distance threshold $d$, and exhaustive search over the integer vectors with at most $d^p$ nonzero coordinates solves SVP in time $(dn)^{O(d^p)}$.
Downey, Fellows, Vardy, and Whittle~\cite{journals/siamcomp/DowneyFVW99} asked whether SVP is $\W[1]$-hard, together with the same question for the parameterized Minimum Distance Problem (MDP) on linear codes, whose natural parameter is also the distance threshold $k$; in their framework, $\W[1]$-hardness refers to the standard deterministic FPT reductions.
For binary codes, MDP is equivalent to the $k$-Even Set problem, whose fixed-parameter tractability remained a central open problem in parameterized complexity for almost two decades~\cite{book/DowneyF99}.

Bhattacharyya, Bonnet, Egri, Ghoshal, Karthik C.\,S., Lin, Manurangsi, and Marx~\cite{journals/BhattacharyyaBE+21} answered both questions under randomized reductions.
They proved that, for every $p>1$, approximating SVP in the $\ell_p$ norm within some constant factor is $\W[1]$-hard, and that approximating MDP over $\F_2$ within any constant factor is $\W[1]$-hard.
Bennett, Cheraghchi, Guruswami, and Ribeiro~\cite{journals/sicomp/BennettCGR24} extended these results to SVP in every finite $\ell_p$ norm and to MDP over every finite field, again under randomized reductions.
Both works asked for deterministic reductions~\cite[Section~8]{journals/BhattacharyyaBE+21}, and Bennett, Cheraghchi, Guruswami, and Ribeiro~\cite{journals/sicomp/BennettCGR24} called finding them a groundbreaking contribution.
Hirahara and Ogitsuka~\cite{conf/random/HiraharaO26} gave one-sided-error reductions, which always map NO instances to NO instances, for MDP within every constant factor and for SVP within every factor below $2^{1/p}$.
They then derandomized these reductions under the assumption that some problem solvable in time $2^{O(n)}$ requires nondeterministic circuits of size $2^{\Omega(n)}$.
For MDP, Guruswami and Ren~\cite{eccc/GuruswamiR26} very recently gave an unconditional deterministic reduction, using an elementary construction based on Reed--Muller codes.
For SVP, no unconditional deterministic reduction was known.

\subsection{Our Results}

We prove the parameterized hardness of SVP under deterministic reductions.
For a constant $\gamma\ge1$, the promise problem $\gamma$-$\GapSVP_p$ asks us to distinguish integer lattices that contain a nonzero vector of $\ell_p$ norm at most $d$ from those in which every nonzero vector has norm greater than $\gamma d$; the parameter is $d$ (see \Cref{def:svp}).

\begin{theorem}
\label{thm:SVP-is-W[1]hard}
For every computable\footnote{A real number $p$ is computable if an algorithm outputs rational approximations of $p$ to any given precision. We use this only to compute the parameters of the reduction.} real $p\ge1$ and every constant $\gamma \in [1, 2^{1/p})$, $\gamma$-$\GapSVP_p$ is $\W[1]$-hard under deterministic $\FPT$ many-one reductions.
\end{theorem}

\Cref{thm:SVP-is-W[1]hard} answers the question of Downey, Fellows, Vardy, and Whittle~\cite{journals/siamcomp/DowneyFVW99} for SVP with deterministic reductions, as well as the question of~\cite{journals/BhattacharyyaBE+21,journals/sicomp/BennettCGR24}: it gives the first deterministic reduction that proves the $\W[1]$-hardness of SVP without any assumption.
Previously, even for exact SVP in the Euclidean norm, $\W[1]$-hardness under deterministic reductions was known only under the circuit lower-bound assumption of~\cite{conf/random/HiraharaO26}.
Its range of approximation factors is the same as that of the one-sided-error reductions of~\cite{conf/random/HiraharaO26} and of the deterministic NP-hardness for general $p$ in~\cite{wan2026}.

Combined with an improved version of the tensor-product amplification of Wan~\cite{wan2026anyconstant}, our approach extends to every constant factor.

\begin{theorem}\label{thm:svp-ext-hardness}
For every computable real $p\ge1$ and every constant $\gamma\ge1$, $\gamma$-$\GapSVP_p$ is $\W[1]$-hard under deterministic $\FPT$ many-one reductions.
\end{theorem}

\Cref{thm:svp-ext-hardness} is the parameterized analogue of the deterministic NP-hardness of~\cite{wan2026anyconstant}.
For $p>1$, its range of approximation factors matches that of the randomized reduction of~\cite{journals/sicomp/BennettCGR24}; for $p=1$, it goes beyond their range, which consists of the factors below $2$.
We prove \Cref{thm:svp-ext-hardness} in \Cref{sec:svp-extend}: we reduce from the Maximum Likelihood Decoding Problem using the construction of \Cref{thm:CVP-to-SVP}, and then amplify the gap by taking tensor powers.

The same method applies to MDP.
For a constant $\gamma\ge1$, the problem $\gamma$-$\GapMDP_q$ asks us to distinguish linear codes over $\F_q$ that contain a nonzero codeword of Hamming weight at most $k$ from those in which every nonzero codeword has weight greater than $\gamma k$ (see \Cref{def:mdp}).

\begin{theorem}
\label{thm:MDP-is-W[1]hard}
For every prime power $q \geq 2$ and every constant $\gamma \geq 1$, $\gamma$-$\GapMDP_q$ is $\W[1]$-hard under deterministic $\FPT$ many-one reductions.
\end{theorem}

Guruswami and Ren~\cite{eccc/GuruswamiR26} proved \Cref{thm:MDP-is-W[1]hard} with a different, elementary construction, starting from the exact nearest codeword problem.
In their reduction, the output threshold grows exponentially with the input threshold $k$.
In ours, it grows linearly: before the final amplification, our reduction maps the approximate nearest codeword problem with threshold $k$ to MDP with threshold $(4p-1)k$ over $\F_p$ (\Cref{thm:NCP-to-MDP-someapprox}).

Guruswami and Ren also mention a reduction with a linear threshold and essentially the same parameters as ours, which GPT-6 discovered using algebraic geometry, but they did not include it because they had not carefully verified its proof~\cite[Footnote~1]{eccc/GuruswamiR26}.
Our main technical contribution is to identify a finite-field version of the implicit function theorem (see \Cref{lem:keep-all-but-one}) as the key idea underlying the algebraic argument, and to give a short proof using standard facts about curves over finite fields.

\subsection{Technical Overview}
Our reductions build directly on those of Hirahara and Ogitsuka~\cite{conf/random/HiraharaO26}.
Hirahara and Ogitsuka build on the reductions of Bennett, Cheraghchi, Guruswami, and Ribeiro~\cite{journals/sicomp/BennettCGR24}, which adapt the approach developed by Khot~\cite{journals/JACM/Kho05} for proving NP-hardness of SVP to the parameterized setting.
Hirahara and Ogitsuka guarantee soundness deterministically, leaving randomness only for completeness.
We make completeness deterministic as well.
Thus, our approach can be viewed as a derandomized version of Khot's approach.
In these approaches, the reduction to SVP starts from its inhomogeneous version, the Closest Vector Problem (CVP).
Similarly, the reduction to the Minimum Distance Problem starts from its inhomogeneous version, the Nearest Codeword Problem (NCP).
We explain the reduction for SVP and then the changes for MDP.

\subsubsection{The Reduction for the Shortest Vector Problem}
\paragraph{Previous randomized approach.}
We briefly recall the reduction of Hirahara and Ogitsuka~\cite{conf/random/HiraharaO26}, on which our reduction is based.
It starts from an instance $(\mathbf B,\vec t,d)$ of CVP with approximation factor $\gamma$, where $\mathbf B\in\mathbb Z^{m\times n}$ is a basis matrix, $\vec t\in\mathbb Z^m$ is a target vector, and $d$ is a positive integer.
This problem asks whether there exists $\vec x\in\mathbb Z^n$ with $\|\mathbf B\vec x-\vec t\|_p\le d$.
In a NO instance, not only $\vec t$ but every nonzero integer multiple of $\vec t$ is far from the lattice: $\|\mathbf B\vec x-\ell\vec t\|_p>\gamma d$ for every $\vec x\in\mathbb Z^n$ and every nonzero integer $\ell$.
The problem remains $\W[1]$-hard with this stronger NO condition.
The reduction uses a \emph{locally dense lattice} with basis $\mathbf A$ and center $\vec s$.
Informally, a locally dense lattice refers to a lattice with many lattice points lying in a ball centered at $\vec s$ whose radius is smaller than the length of a shortest nonzero lattice vector.
In~\cite{journals/sicomp/BennettCGR24,conf/random/HiraharaO26}, the basis $\mathbf A$ is constructed deterministically so that the lattice it generates has minimum norm greater than $\gamma d$.
The center $\vec s$ is chosen randomly to obtain many short vectors $\mathbf A\vec y-\vec s$.

For a suitable prime $q>\gamma d$ and an integer $\beta>\gamma d$,
the output lattice consists of vectors of the form
\[
 \begin{pmatrix}
  \mathbf B&\mathbf0&-\vec t&\mathbf0\\
  \mathbf0&\mathbf A&-\vec s&\mathbf0\\
  \mathbf0&\mathbf0&1&\mathbf0\\
  \beta\mathbf P\mathbf B&
  \beta\mathbf R\mathbf A&
  -\beta(\mathbf P\vec t+\mathbf R\vec s)&
  \beta q\mathbf I
 \end{pmatrix}
 \begin{pmatrix}
  \vec x\\
  \vec y\\
  \ell\\
  \vec z
 \end{pmatrix}
 =
 \begin{pmatrix}
  \mathbf B\vec x-\ell\vec t\\
  \mathbf A\vec y-\ell\vec s\\
  \ell\\
  \beta\bigl(
   \mathbf P(\mathbf B\vec x-\ell\vec t)
   +\mathbf R(\mathbf A\vec y-\ell\vec s)
   +q\vec z
  \bigr)
 \end{pmatrix},
\]
where $\vec x,\vec y,\vec z$ are integer vectors and $\ell$ is an integer.
Here $\mathbf P$ is a deterministically constructed parity-check matrix that maps every nonzero integer vector of $\ell_p$ norm at most $\gamma d$ to a nonzero vector modulo $q$.
The matrix $\mathbf R$ is chosen randomly, and both matrices are represented by integer entries.
The coefficient $\ell$ multiplies both targets, and the last block vanishes for some integer vector $\vec z$ if and only if
\[
 \mathbf P(\mathbf B\vec x-\ell\vec t)
 +\mathbf R(\mathbf A\vec y-\ell\vec s)
 =\vec0\pmod q.
\]

To prove completeness, let $\vec{x}$ satisfy $\|\mathbf{B}\vec{x}-\vec{t}\|_p\le d$.
By the local density, with high probability over the random choice of $\vec{s}$, there are many integer vectors $\vec{y}$ for which $\mathbf{A}\vec{y}-\vec{s}$ is short.
Since there are many such short vectors, with high probability over the random choice of $\mathbf{R}$, one of them is mapped to the required vector, that is,
\begin{equation}\label{eq:linearmap-intro}
 \mathbf{R}(\mathbf{A}\vec{y}-\vec{s})=-\mathbf{P}(\mathbf{B}\vec{x}-\vec{t})\pmod q.
\end{equation}
Setting $\ell=1$ and choosing $\vec z$ so that the last block vanishes gives a short nonzero output vector.

Soundness holds for every choice of $\vec s$ and $\mathbf R$~\cite[Theorem 3.9]{conf/random/HiraharaO26}, so randomness is needed only for completeness.

\paragraph{Our approach: deterministic reduction with locally dense lattices.}
Our reduction is obtained from the reduction of Hirahara and Ogitsuka~\cite{conf/random/HiraharaO26} by a specific choice of $\mathbf{P}$ and $\mathbf{R}$ and by removing the columns $\beta q\mathbf{I}$.
Let $\Lambda$ be the lattice generated by $\mathbf{A}$, and fix an ordered list $S$ of its coordinates.
We write $\vec{v}|_S$ for the restriction of a vector $\vec{v}$ to the coordinates in $S$, in the specified order, and $\mathbf{A}_S$ for the corresponding rows of $\mathbf{A}$.
We take $\mathbf{P}=\mathbf{I}$ and let $\mathbf{R}$ be the negative of the restriction map to $S$, that is, $\mathbf{R}\vec{v}=-\vec{v}|_S$.
Thus $\mathbf{R}$ is no longer random, and we also choose the center $\vec{s}$ deterministically, with $\vec{s}|_S=\vec{0}$.
With these choices, \Cref{eq:linearmap-intro} becomes $(\mathbf{A}\vec{y}-\vec{s})|_S=\mathbf{B}\vec{x}-\vec{t}\pmod q$.
Since we remove the columns $\beta q\mathbf{I}$, which account for the reduction modulo $q$, we require this to hold as an equality of integer vectors:
\[
 (\mathbf{A}\vec{y}-\vec{s})|_S=\mathbf{B}\vec{x}-\vec{t}.
\]
To satisfy this for every YES instance without randomness, every short vector in $\Z^m$ must be the restriction to $S$ of some short vector in $\Lambda-\vec{s}$.
In~\cite{conf/random/HiraharaO26}, a locally dense lattice consists only of the basis $\mathbf{A}$ and the center $\vec{s}$, and the random matrix $\mathbf{R}$ is responsible for satisfying~\Cref{eq:linearmap-intro}.
We instead include the list $S$ in our locally dense lattice and construct $\vec{s}$ and $S$ deterministically so that the restriction to $S$ plays this role.

Accordingly, our locally dense lattice consists of a lattice $\Lambda$ whose minimum $\ell_p$ norm is greater than $\gamma d$, a center $\vec{s}$, and an ordered list $S$ of coordinates with $\vec{s}|_S=\vec{0}$, such that every short vector in $\Z^m$ is the restriction to $S$ of a short vector in $\Lambda-\vec{s}$:
\begin{equation}\label{eq:intro-completion}
 \parbox{0.85\textwidth}{For every $\vec{u}\in\Z^m$ with $\|\vec{u}\|_p\le d$, some $\vec{w}\in\Lambda-\vec{s}$ with $\|\vec{w}\|_p^p\le W$ satisfies $\vec{w}|_S=\vec{u}$.}
\end{equation}
Condition~\Cref{eq:intro-completion} is our counterpart of~\Cref{eq:linearmap-intro}.
Taking $\vec{u}=\mathbf{B}\vec{x}-\vec{t}$, it guarantees the equality above for every $\vec{x}$ with $\|\mathbf{B}\vec{x}-\vec{t}\|_p\le d$, without any randomness.

For an integer $\beta>\gamma d$, given a locally dense lattice $(\Lambda,\vec{s},S)$ satisfying \Cref{eq:intro-completion} as advice, the reduction \emph{deterministically} outputs a lattice consisting of vectors of the form
\[
 \begin{pmatrix}
  \mathbf{B}&\mathbf{0}&-\vec{t}\\
  \mathbf{0}&\mathbf{A}&-\vec{s}\\
  \mathbf{0}&\mathbf{0}&1\\
  \beta\mathbf{B}&-\beta\mathbf{A}_S&-\beta\vec{t}
 \end{pmatrix}
 \begin{pmatrix}
  \vec{x}\\
  \vec{y}\\
  \ell
 \end{pmatrix}
 =
 \begin{pmatrix}
  \mathbf{B}\vec{x}-\ell\vec{t}\\
  \mathbf{A}\vec{y}-\ell\vec{s}\\
  \ell\\
  \beta(\mathbf{B}\vec{x}-\ell\vec{t}-\mathbf{A}_S\vec{y})
 \end{pmatrix},
\]
where $\vec{x},\vec{y}$ are integer vectors and $\ell$ is an integer.

To prove completeness, consider a YES instance of CVP and let $\vec x$ satisfy $\|\mathbf B\vec x-\vec t\|_p\le d$.
Apply \Cref{eq:intro-completion} with $\vec{u}\gets \mathbf{B}\vec{x}-\vec{t}$.
The resulting vector $\vec{w}$ can be written as $\mathbf{A}\vec{y}-\vec{s}$ for an integer vector $\vec{y}$.
Since $\vec{s}|_S=\vec{0}$, the equality $\vec{w}|_S=\mathbf{B}\vec{x}-\vec{t}$ gives $\mathbf{A}_S\vec{y}=\mathbf{B}\vec{x}-\vec{t}$.
With $\ell=1$, the last block vanishes.
The output vector is then nonzero because its third block is $\ell=1$, and the $p$th power of its norm is $\|\mathbf B\vec x-\vec t\|_p^p+\|\mathbf A\vec y-\vec s\|_p^p+1\le d^p+W+1$.

Soundness does not require \Cref{eq:intro-completion}; it uses only the minimum norm bound for $\Lambda$ and the inequality $\beta>\gamma d$.
Consider a NO instance of CVP and an output vector of norm at most $\gamma d$.
The promise of CVP forces $\ell=0$, since otherwise $\|\mathbf B\vec x-\ell\vec t\|_p>\gamma d$.
Then $\mathbf A\vec y=\vec0$, because it belongs to $\Lambda$, has norm at most $\gamma d$, and every nonzero vector in $\Lambda$ has norm greater than $\gamma d$.
The last block is now $\beta\mathbf B\vec x$.
Since $\mathbf B\vec x$ is an integer vector, this block is either zero or has norm at least $\beta>\gamma d$.
Thus $\mathbf B\vec x=\vec0$, so every block is zero, proving soundness.
It remains to construct $(\Lambda,\vec s,S)$ deterministically in time $f(d)m^{O(1)}$, with $W$ bounded by a function of $d$ alone.

\paragraph{Deterministic construction of locally dense lattices.}
We provide a technical overview of the \emph{deterministic} construction of locally dense lattices required above.
For an integer $k\ge2$ and a sufficiently large prime $q>2k$, we use the Reed--Solomon lattice of Bennett and Peikert~\cite{conf/random/BennettP23},
\begin{equation}\label{eq:intro-rs-lattice}
 \Lambda=
 \Bigl\{
  \vec{v}\in\Z^{\F_q}:
  \sum_{z\in\F_q}v_z z^j=0\pmod q
  \text{ for every }0\le j<k
 \Bigr\}.
\end{equation}
Field elements label the coordinates, while the vector entries are integers.
It was shown in \cite{conf/random/BennettP23} that the minimum distance of $\Lambda$ in the $\ell_p$ norm is at least $(2k)^{1/p}$.

We construct the center $\vec s$ and the ordered list $S$ of $m$ coordinates satisfying~\Cref{eq:intro-completion} with $W=k-1+2d^p$.
Fix distinct points $y_1,\ldots,y_{k-1},x_1,\ldots,x_m\in\F_q$.
Set $S=(x_1,\ldots,x_m)$ and $\vec{s}=-\sum_{\ell=1}^{k-1}\vec{e}_{y_\ell}$, where $\vec{e}_\xi$ is the unit vector at coordinate $\xi\in\F_q$.
For clarity, we reorder the coordinates in each vector display below to show where the entries are placed.
The center has the form
\[
 \vec{s}=
 \bigl(
  \underbrace{0,\ldots,0}_{x_1,\ldots,x_m},
  \underbrace{-1,\ldots,-1}_{y_1,\ldots,y_{k-1}},
  \underbrace{0,\ldots,0}_{\text{remaining coordinates}}
 \bigr)^{T}.
\]
In particular, $\vec{s}|_S=\vec{0}$.

For $\vec u=\vec0$, we may take $\vec w=-\vec s$.
Then $\vec w\in\Lambda-\vec s$, $\vec w|_S=\vec u$, and $\|\vec w\|_p^p=k-1\le W$, as required in~\Cref{eq:intro-completion}.
Now fix a nonzero $\vec{u}$ with $\|\vec{u}\|_p\le d$, and let $I=\{i:u_i\ne0\}$.
We seek a vector of the form
\[
 \vec{w}
 =\sum_{\ell=1}^{k-1}\vec{e}_{z_\ell}
  +\sum_{i\in I}u_i
   \bigl(\vec{e}_{x_i}-\vec{e}_{x_i+\tau}\bigr),
 \qquad
 z_1,\ldots,z_{k-1},\tau\in\F_q.
\]
The second sum places $\vec{u}$ on $S$ together with a copy of $-\vec{u}$ shifted by $\tau$, and we seek $\tau$ and the positions $\{z_\ell\}_{1\le \ell <k}$ so that $\vec{s}+\vec{w}\in\Lambda$.
The required restriction and norm bound follow if the positions $z_\ell$ and $x_i+\tau$ are distinct and outside $S$.
In this case, $\vec{w}$ has the form
\[
 \vec{w}=
 \bigl(
  \underbrace{u_1,\ldots,u_m}_{x_1,\ldots,x_m},
  \underbrace{1,\ldots,1}_{z_1,\ldots,z_{k-1}},
  \underbrace{(-u_i)_{i\in I}}_{(x_i+\tau)_{i\in I}},
  \underbrace{0,\ldots,0}_{\text{remaining coordinates}}
 \bigr)^T.
\]
Consequently,
\[
 \vec{w}|_S=\vec{u},
 \qquad
 \|\vec{w}\|_p^p
 =k-1+2\|\vec{u}\|_p^p
 \le k-1+2d^p.
\]
For vectors of this form, the condition $\vec{s}+\vec{w}\in\Lambda$ is equivalent to $k-1$ polynomial equations in the coordinate labels.
The condition in \Cref{eq:intro-rs-lattice} for $j=0$ holds automatically because the entries of $\vec{s}+\vec{w}$ sum to zero.
The remaining conditions are $f_j(z_1,\ldots,z_{k-1},\tau)=0$ for $1\le j<k$, where
\[
 \begin{aligned}
 f_j(Z_1,\ldots,Z_{k-1},Y)
 &=
 \sum_{\ell=1}^{k-1}(Z_\ell^j-y_\ell^j)
 -
 \sum_{i\in I}u_i\bigl((x_i+Y)^j-x_i^j\bigr).
 \end{aligned}
\]
These equations have a common zero $\xi=(y_1,\ldots,y_{k-1},0)$.
At this point, the entries $u_i$ and $-u_i$ occupy the same coordinates and cancel, leaving $\vec{w}=-\vec{s}$ and $\vec{w}|_S=\vec{0}$.
Our goal is to find another common zero for which the field elements $z_1,\ldots,z_{k-1}$ and $x_i+\tau$ ($i\in I$) are all distinct and lie outside $S$.
We express these requirements by one more polynomial $g$, which is nonzero only when they hold. 
Let $E=\{x_b-x_i:b\in[m],\,i\in I\}$ and define
\[
 \begin{aligned}
 g(Z_1,\ldots,Z_{k-1},Y)
 &=
 Y\prod_{e\in E\setminus\{0\}}(Y-e)
 \prod_{\ell=1}^{k-1}\prod_{b=1}^{m}(Z_\ell-x_b)
 \cdot
 \prod_{1\le\ell<\ell'<k}(Z_\ell-Z_{\ell'})
 \prod_{\ell=1}^{k-1}\prod_{i\in I}(Z_\ell-x_i-Y).
 \end{aligned}
\]
It is straightforward to verify that $g(\xi)=0$. If $g(z_1,\ldots,z_{k-1},\tau)\ne0$, the factors involving only $Y$ ensure that $x_i+\tau\notin S$ for every $i\in I$.
The remaining products ensure, respectively, that the $z_\ell$ lie outside $S$, that they are distinct from one another, and that none equals $x_i+\tau$ for any $i\in I$.
The elements $x_i+\tau$ are distinct from one another because the $x_i$ are distinct.
Over the real numbers, the implicit function theorem would let us move $\tau$ slightly away from $0$ and adjust $z_1,\ldots,z_{k-1}$ so that the equations still hold, provided that the Jacobian matrix with respect to $Z_1,\ldots,Z_{k-1}$ is invertible at $\xi$.
For small nonzero $\tau$, the conditions above would then hold as well.
A finite field has no notion of a small change, so we instead use the following finite-field version of the implicit function theorem.

\begin{lemma}[Informal; see \Cref{lem:keep-all-but-one}]
\label{lem:intro-keep-all-but-one}
Let $f_1,\ldots,f_n\in\F_Q[X_1,\ldots,X_n]$ have a common zero $\xi\in\F_Q^n$ at which their Jacobian matrix is invertible.
If $Q$ is large compared with the degrees of $f_1,\ldots,f_n$, then there exists $x\in\F_Q^n$ with $f_1(x)=\cdots=f_{n-1}(x)=0$ and $f_n(x)\ne0$.
\end{lemma}

We now verify the Jacobian condition in \Cref{lem:intro-keep-all-but-one}.
The Jacobian matrix of $f_1,\ldots,f_{k-1}$ with respect to $Z_1,\ldots,Z_{k-1}$ at $\xi$ is
\[
 (jy_\ell^{j-1})_{1\le j,\ell<k},
\]
which is invertible because the $y_\ell$ are distinct and $q>k$.
Only the factor $Y$ in $g$ vanishes at $\xi$.
Thus $g(\xi)=0$, $\partial g/\partial Y$ is nonzero at $\xi$, and every $\partial g/\partial Z_\ell$ vanishes there.
Together, these facts show that the Jacobian of $f_1,\ldots,f_{k-1},g$ with respect to $Z_1,\ldots,Z_{k-1},Y$ is invertible at $\xi$.

For sufficiently large $q$, applying \Cref{lem:intro-keep-all-but-one} to $f_1,\ldots,f_{k-1},g$ gives a common zero $x=(z_1,\ldots,z_{k-1},\tau)$ of the $f_j$'s at which $g\ne0$.
The resulting vector $\vec w$ belongs to $\Lambda-\vec s$ and satisfies
\[
 \vec w|_S=\vec u,
 \qquad
 \|\vec w\|_p^p=k-1+2\|\vec u\|_p^p
 \le k-1+2d^p,
\]
as required in~\Cref{eq:intro-completion}.

Finally, we explain why the construction takes FPT time.
Once the prime $q$ is chosen, we form the $k\times q$ matrix $(z^j)_{0\le j<k,\,z\in\F_q}$ that defines $\Lambda$.
Gaussian elimination over $\F_q$ reduces the defining congruences to a form from which we can construct an integer basis of $\Lambda$.
We also choose $m+k-1$ distinct field elements and use them to define $\vec{s}$ and $S$ as above.
These computations take time polynomial in $q$.
For sufficiently large $\gamma$, we choose $k$ as a function of $d$ just above $(\gamma d)^p/2$, so that $k>d^p$ and $(2k)^{1/p}>\gamma d$.
The resulting approximation factor $\gamma d/(k+3d^p)^{1/p}$ approaches $2^{1/p}$ as $\gamma$ grows.
For every $\vec{u}\in\Z^m$ with $\|\vec{u}\|_p\le d$, its support $I=\{i\in[m]:u_i\ne0\}$ has size $|I|\le d^p<k$.
The degrees of the $f_j$ are $1,\ldots,k-1$, while $\deg g=O(km+k^2)$.
By \Cref{lem:intro-keep-all-but-one}, the required field size is therefore at most a function of $k$ times $m$.
We can thus choose a prime $q\le f(k)m$ that works for every such $\vec{u}$.
Finding such a prime and carrying out the construction take time polynomial in $q$, giving a total running time of $f(d)m^{O(1)}$.

Wan~\cite{wan2026} also constructs locally dense lattices from Reed--Solomon codes deterministically, and counts the points of higher-dimensional varieties using Deligne's theorem.
Our lemma needs only the Weil bound for curves~\cite{Schmidt1976,CafureMatera2006}.
In return, it requires the field size to exceed a fixed power of the product of the degrees of $f_1,\ldots,f_{n-1}$.
The parameterized setting can afford this, because these degrees depend only on the parameter.

\subsubsection{The Reduction for the Minimum Distance Problem}
We apply an idea similar to that used for lattices to codes to establish the $\W[1]$-hardness of MDP.
More specifically, our reduction builds on that of Hirahara and Ogitsuka~\cite{conf/random/HiraharaO26}, and we deterministically construct a \emph{locally dense code}, the coding analogue of a locally dense lattice.
Informally, many codewords lie in a Hamming ball centered at $\vec{s}$ whose radius is smaller than the minimum distance of the code.

To reduce NCP to MDP, we use selected coordinates of the locally dense code to represent the difference between an input codeword and the target.
Given an NCP instance $(\mathbf{G},\vec{t},k)$ with $m$ coordinates, we construct a code generated by $\mathbf{A}$, a center $\vec{s}$, and an ordered list $S$ of $m$ coordinates, with $\vec{s}|_S=\vec{0}$.
For every $\vec{x}$ satisfying $\|\mathbf{G}\vec{x}-\vec{t}\|_0\le k$, there must be a vector $\mathbf{A}\vec{y}-\vec{s}$ that is nonzero, has Hamming weight bounded by a fixed multiple of $k$, and satisfies
\[
 (\mathbf{A}\vec{y}-\vec{s})|_S=\mathbf{G}\vec{x}-\vec{t}.
\]
Since $\vec{s}|_S=\vec{0}$, this equality gives $\mathbf{G}\vec{x}-\vec{t}-\mathbf{A}_S\vec{y}=\vec{0}$.
We require $\mathbf{A}\vec{y}-\vec{s}$ to be nonzero even when $\mathbf{G}\vec{x}=\vec{t}$, so that the output contains a nonzero codeword in this case as well.

The reduction replaces multiplication by a large integer with repetition of coordinates.
Multiplication by a nonzero field element preserves Hamming weight, whereas repeating each coordinate multiplies the weight by the number of repetitions.
In a YES instance, the repeated coordinates are copies of the entries of $\mathbf{G}\vec{x}-\vec{t}-\mathbf{A}_S\vec{y}$ and are therefore zero.
We choose the number of repetitions so that any nonzero entry in these coordinates makes the output weight exceed the NO threshold.
A nonzero coefficient of the target can be divided out without changing Hamming weight, so the NO condition of NCP applies to that case.

\paragraph{Constructing the locally dense codes.}
We construct the locally dense code from an extended BCH code over a fixed prime field $\F_p$.
Let $K$ be a sufficiently large extension field of $\F_p$, and let $D$ be the chosen distance bound.
Codeword entries belong to $\F_p$, while coordinates are indexed by $K$.
We define
\[
 \C=\left\{\vec c\in\F_p^K:
    \sum_{z\in K}c_z=0,\quad
    \sum_{z\in K}c_z z^j=0
    \quad(1\le j<D,\ p\nmid j)
 \right\}.
\]
Here we omit the equations for positive exponents divisible by $p$.
Since $c_z^p=c_z$ for every $c_z\in\F_p$, we have
\[
 \left(\sum_{z\in K}c_z z^j\right)^p
 =
 \sum_{z\in K}c_z z^{pj}.
\]
Applying this identity repeatedly gives all the omitted equations.
Thus, the code is the same as the one defined by imposing the moment equations for every exponent $0\le j<D$, and its minimum distance is unchanged.

We use the same expression for $\vec{w}$ as in the lattice construction, now with coefficients in the prime field and one added unit vector for each remaining positive exponent.
The coordinates of $\vec{s}+\vec{w}$ sum to zero, so only the moment equations with positive exponents remain to be satisfied.
Omitting the redundant equations reduces the number of added unit vectors and hence the upper bound on the Hamming weight of $\vec{w}$.
Choosing the BCH distance sufficiently large compared with $k$ then gives a constant gap between the YES and NO thresholds.

We choose the support of $\vec{s}$ so that the Jacobian with respect to the variables $Z_\ell$ is invertible at the known solution.
As in the lattice construction, each $Z_\ell$ specifies the coordinate at which an added unit vector is placed.
For the remaining positive exponents $j$, the derivatives $jZ^{j-1}$ are linearly independent polynomials.
Evaluating them over a sufficiently large extension field gives a matrix of full row rank.
We select independent columns by Gaussian elimination and use their indices as the support of $\vec{s}$.

The polynomial $g$ only needs to prevent the added terms from changing the entries on $S$.
We choose it so that $g\ne0$ forces the coordinates occupied by the translated negative of $\mathbf{G}\vec{x}-\vec{t}$ and by the added unit vectors to lie outside $S$.
These coordinates may coincide with one another, since such coincidences cannot increase Hamming weight.
For a nonzero difference $\mathbf{G}\vec{x}-\vec{t}$, the lemma above therefore guarantees a vector $\vec{w}$ of Hamming weight at most a fixed multiple of $k$ such that $\vec{s}+\vec{w}$ belongs to the code and $\vec{w}|_S=\mathbf{G}\vec{x}-\vec{t}$.
When the difference is zero, we take $\vec{w}=-\vec{s}$.
As before, the field size bound allows us to construct $\mathbf{A}$, $\vec{s}$, and $S$ in deterministic FPT time.

\paragraph{All fixed finite fields and any constant approximation factor.}
We extend the result to every fixed finite field and every constant approximation factor by extending the coefficient field and taking tensor powers.
Viewing the same generator matrix over an extension field preserves its minimum distance.
The minimum distance of a tensor product of codes is the product of their minimum distances, so a constant number of tensor factors amplifies the gap to any prescribed constant.

\subsection{Related Work}
Ajtai--Micciancio's approach uses locally dense lattices to prove the classical NP-hardness of SVP.
The notion of a locally dense lattice used in our reductions is a variant of the one used in this approach.
However, the existing constructions in this approach do not directly apply to the parameterized setting considered here \cite[Section~1.2]{conf/random/HiraharaO26}.
This approach originated in the work of Ajtai~\cite{conf/stoc/Ajtai98}, whose construction uses a variant of the Schnorr--Adleman prime number lattice.
Micciancio~\cite{journals/sicomp/Micciancio01} used a simpler variant of this lattice to prove NP-hardness of approximating SVP under randomized reductions with one-sided error.
Micciancio~\cite{journals/toc/Micciancio12} later constructed locally dense lattices by applying ``Construction D''~\cite{Conway99_book} to a nested sequence of extended binary BCH codes and obtained such reductions for every constant approximation factor in the Euclidean norm.
Bennett and Peikert~\cite{conf/random/BennettP23} gave a simpler construction by applying ``Construction A''~\cite{Conway99_book} to Reed--Solomon codes.
Recently, Wan~\cite{wan2026,wan2026anyconstant} derandomized this approach by deterministically constructing a suitable center and projection for these Reed--Solomon lattices.

Dumer, Micciancio, and Sudan~\cite{journals/DumerMS03} adapted Ajtai--Micciancio's approach to linear codes, using locally dense codes to prove NP-hardness of approximating MDP under randomized reductions.
Cheng and Wan~\cite{journals/IEEE/ChengW12} derandomized this reduction using Reed--Solomon codes concatenated with Hadamard codes.
They explicitly constructed a center and a coordinate projection satisfying the properties required by the reduction.
Micciancio~\cite{conf/ccc/Micciancio14} later gave a general deterministic construction that transforms binary codes with sufficiently large relative minimum distance into locally dense binary codes.
His construction uses tensor products and produces an explicit center and coordinate projection, providing another way to derandomize the reduction of Dumer, Micciancio, and Sudan.

Khot~\cite{journals/JACM/Kho05} developed another approach to proving NP-hardness of approximating SVP, combining locally dense lattices with random sparsification.
His construction uses lattices obtained from binary BCH codes.
Haviv and Regev~\cite{journals/toc/HR12} subsequently showed that standard tensor products of the lattices arising in Khot's reduction can amplify the approximation gap.

Variants of Khot's approach have also been used to establish fine-grained hardness of SVP.
Aggarwal and Stephens-Davidowitz~\cite{conf/stoc/AggarwalS18} showed that, under randomized Gap-ETH, for every fixed $p>2$, there is a constant approximation factor greater than one for which SVP in the $\ell_p$ norm admits no algorithm whose running time is subexponential in the lattice rank.
Aggarwal, Gupta, Morolia, and Zhang~\cite{aggarwal2026mindgapsvphardness} later obtained analogous lower bounds under randomized ETH.
\footnote{Recent preprints by OpenAI~\cite{OpenAI2026ETH} and Anthropic~\cite{Claude2026ETH} give deterministic algorithms for solving $3$-SAT in subexponential time.
The reductions used to prove the SVP hardness results do not rely on the stated hypotheses.
These hypotheses are needed only to derive the lower bounds on running time.}
The locally dense lattices used in these reductions still have many lattice points close to a common center, but the required local density condition differs from those discussed above.
Specifically, the number of lattice points in a ball around the center must be exponentially larger than the number in a slightly larger ball around the origin.
The radius of the ball around the center need not be smaller than the length of a shortest nonzero lattice vector.

\subsection{Open Problems}
\Cref{thm:SVP-is-W[1]hard,thm:svp-ext-hardness} concern constant approximation factors.
In the classical setting, Hair and Sahai~\cite{HairSahai2026Polynomial} proved that, for every $p>2$, SVP in the $\ell_p$ norm is NP-hard to approximate within a polynomial factor under deterministic reductions.
A natural question is whether $\gamma$-$\GapSVP_p$ remains $\W[1]$-hard when $\gamma$ grows polynomially with the rank of the lattice.

Another open problem is to make \Cref{lem:intro-keep-all-but-one} constructive, that is, to give an efficient algorithm that finds a point $x$ with $f_1(x)=\cdots=f_{n-1}(x)=0$ and $f_n(x)\ne0$.

\paragraph{Organization.}
\Cref{sec:prelim} gives the preliminaries and proves the key \Cref{lem:keep-all-but-one} in \Cref{sec:curve-lemma}.
\Cref{sec:svp} proves \Cref{thm:SVP-is-W[1]hard}, our main result on SVP, and \Cref{sec:mdp} proves \Cref{thm:MDP-is-W[1]hard}, our main result on MDP. 
\Cref{sec:svp-extend} extends the result on SVP to every constant factor and proves \Cref{thm:svp-ext-hardness}.

\section{Preliminaries}\label{sec:prelim}

\subsection{Notation}
For a positive integer $n$, write $[n]=\{1,\ldots,n\}$.
We regard vectors as column vectors, even when their coordinates are written as a tuple.
For a matrix $\mathbf A$, we write $A_{ij}$ for its entry in row $i$ and column $j$.
The notation $(a_{ij})_{i\in I,\,j\in J}$ denotes the matrix with rows indexed by $I$, columns indexed by $J$, and entry $a_{ij}$ in row $i$ and column $j$.
We write $\vec0$ for a zero vector and $\mathbf0$ for a zero matrix, with dimensions determined by context.
The matrix $\mathbf I_n$ is the $n\times n$ identity matrix.
In a block matrix, matrices and column vectors are placed side by side or one above another, with matching dimensions.
For a finite set $T$ and a field $K$, $K^T$ denotes vectors with entries in $K$ indexed by $T$; $\Z^T$ denotes integer vectors indexed by $T$.
For $z\in T$, the vector $\vec e_z\in K^T$ has entry $1$ at $z$ and $0$ elsewhere; the same notation denotes the corresponding vector in $\Z^T$.
The support of $\vec v$ is $\{z:v_z\ne0\}$, and its cardinality $\|\vec v\|_0$ is its Hamming weight.
For a real vector, $\|\vec v\|_p=(\sum_z|v_z|^p)^{1/p}$ for $1\le p<\infty$, and $\|\vec v\|_\infty=\max_z|v_z|$.

For an ordered list $S=(x_1,\ldots,x_m)$ of distinct coordinates, put $\vec v|_S=(v_{x_1},\ldots,v_{x_m})$.
For a set of vectors $V$, write $V-\vec s=\{\vec v-\vec s:\vec v\in V\}$.
When $V$ is a linear code or a lattice, this translate is a coset.
The Kronecker product $\mathbf G\otimes\mathbf A$ is the block matrix whose $(i,j)$ block is $G_{ij}\mathbf A$; $\mathbf G^{\otimes\ell}$ denotes the Kronecker product of $\ell$ copies of $\mathbf G$.

A real number is computable if an algorithm, given $r\in\Z_{>0}$, returns a rational approximation with error at most $2^{-r}$.

\subsection{Finite fields and polynomials}
For a prime power $q$, $\F_q$ denotes the field with $q$ elements.
For prime $p$, $\F_p=\Z/p\Z$.
A field extension $K/F$ is a field $K$ containing $F$ as a subfield.
Its degree is the dimension of $K$ as a vector space over $F$.
An $F$-basis of $K$ expresses each element of $K$ uniquely as an $F$-linear combination of the basis elements.
The field $\F_{q^s}$ is an extension of $\F_q$ of degree $s$.

For a field $K$ and variables $X_1,\ldots,X_r$, the ring $K[X_1,\ldots,X_r]$ consists of polynomials in these variables with coefficients in $K$.
For a tuple of variables $X=(X_1,\ldots,X_r)$, write $K[X]=K[X_1,\ldots,X_r]$ and $f(X)=f(X_1,\ldots,X_r)$ for $f\in K[X]$.
For a nonzero polynomial $f\in K[X_1,\ldots,X_r]$, its total degree $\deg f$ is the largest sum of exponents among its monomials with nonzero coefficient.
For $i\in[r]$, its degree in $X_i$, denoted by $\deg_{X_i}f$, is the largest exponent of $X_i$ among those monomials.
We set $\deg 0=\deg_{X_i}0=-\infty$.
For elements $a_1,\ldots,a_r\in K$, the value $f(a_1,\ldots,a_r)$ is obtained by substituting $a_i$ for $X_i$ for every $i\in[r]$.
A common zero of several polynomials is a tuple at which all of them vanish.
An affine linear polynomial in $K[X_1,\ldots,X_r]$ has the form $a_0+\sum_{i=1}^r a_iX_i$, where $a_0,\ldots,a_r\in K$.
It is nonconstant if at least one $a_i$ with $i\ge1$ is nonzero.

For a polynomial $f\in K[X_1,\ldots,X_r]$, the formal partial derivative $\partial f/\partial X_i$ is obtained term by term, treating the other variables as constants and using $\partial X_i^a/\partial X_i=aX_i^{a-1}$ for $a\ge1$ and $\partial 1/\partial X_i=0$.
The coefficient $a$ is interpreted in $K$.
For polynomials $f_1,\ldots,f_r\in K[X_1,\ldots,X_n]$ and a point $\xi\in K^n$, their Jacobian matrix at $\xi$ is the $r\times n$ matrix
\[
  \left(\frac{\partial f_i}{\partial X_j}(\xi)\right)_{i\in[r],\,j\in[n]}.
\]
For distinct $a_1,\ldots,a_r\in K$, the Vandermonde matrix $(a_j^{i-1})_{i,j\in[r]}$ has determinant $\prod_{i<j}(a_j-a_i)\ne0$.

A nonzero univariate polynomial is monic if its leading coefficient is $1$.
A nonconstant polynomial $f\in K[X_1,\ldots,X_r]$ is irreducible over $K$ if it cannot be written as a product of two nonconstant polynomials in $K[X_1,\ldots,X_r]$.
For a nonzero polynomial $f\in K[X]$ and a root $a\in K$, the multiplicity of $a$ is the largest integer $e\ge1$ such that $(X-a)^e$ divides $f$.
For a nonconstant polynomial $g\in\F_q[X]$, the quotient ring $\F_q[X]/(g)$ consists of polynomials modulo $g$, with addition and multiplication modulo $g$.
If $g$ is irreducible of degree $s$, this quotient is a field with $q^s$ elements, represented by polynomials of degree less than $s$.

\subsection{Lattice and Coding Problems}
We recall the basic definitions of parameterized promise problems.
For details on parameterized complexity, see \cite{book/DowneyF99,journals/tcs/DF95b,journals/sicomp/DF95a}.
\begin{definition}[Parameterized promise problems]
A \emph{parameterized promise problem} is a pair of disjoint languages $\Pi=(\Pi_{\mathrm{YES}},\Pi_{\mathrm{NO}})$ over $\Sigma^* \times \mathbb{N}$.
An instance of $\Pi$ is a pair $(x,k)\in \Sigma^* \times \mathbb{N}$, where $k$ is called the \emph{parameter}.
The set $\Pi_{\mathrm{YES}} \cup \Pi_{\mathrm{NO}}$ is called the \emph{promise}.
\end{definition}

Given a matrix $\mathbf{B}\in\mathbb{Z}^{m\times n}$, the lattice generated by its columns is $\mathcal{L}=\mathcal{L}(\mathbf{B}):=\{\mathbf{B}\vec{z}:\vec{z}\in\mathbb{Z}^n\}\subseteq\mathbb{Z}^m$.
When the columns of $\mathbf{B}$ are linearly independent, $\mathbf{B}$ is referred to as a lattice basis for $\mathcal{L}(\mathbf{B})$.
For $p\in[1,\infty]$, the length of a shortest nonzero lattice vector, measured in the $\ell_p$ norm, is given by $\lambda_1^{(p)}(\mathcal{L}):=\min_{\vec{v}\in\mathcal{L}\setminus\{\mathbf{0}\}}\|\vec{v}\|_p$.
Given a target vector $\vec{t}\in\mathbb{R}^m$, its distance from the lattice is denoted by $\operatorname{dist}_p(\mathcal{L},\vec{t}):=\min_{\vec{v}\in\mathcal{L}}\|\vec{v}-\vec{t}\|_p$.

The Shortest Vector Problem (SVP) is the problem of finding a shortest nonzero vector in a lattice, given a basis matrix of the lattice as input.
Here, we consider an approximate version of the parameterized decision problem in which a parameter is given as an additional input, and the goal is to determine whether the length of a shortest nonzero lattice vector is at most the parameter.
\begin{definition}[$\gamma$-$\GapSVP_p$, \cite{journals/GoldreichG00}]
\label{def:svp}
For $p \geq 1$ and $\gamma \geq 1$, the parameterized promise problem $\gamma$-approximate Shortest Vector Problem ($\gamma$-$\GapSVP_p$) is defined as follows.
Instances are pairs $(\mathbf{B}, d)$, where $\mathbf{B} \in \mathbb{Z}^{m\times n}$ is a lattice basis of $\mathcal{L}(\mathbf{B})$, and $d \in \mathbb{Z}_{>0}$ is a distance parameter such that
\begin{enumerate}
    \item $(\mathbf{B}, d)$ is a YES instance if $\lambda^{(p)}_1(\mathcal{L}(\mathbf{B}))\leq d$;
    \item $(\mathbf{B}, d)$ is a NO instance if $\lambda^{(p)}_1(\mathcal{L}(\mathbf{B}))> \gamma d$.
\end{enumerate}
The parameter of interest is $d$.
\end{definition}

To establish a reduction to SVP, we next define its inhomogeneous version, the Closest Vector Problem (CVP).
In this definition, we require that, for NO instances, every nonzero integer multiple of the target vector is far from every lattice vector.
\begin{definition}[$\gamma$-$\GapCVP_p$, \cite{journals/sicomp/BennettCGR24}]
\label{def:cvp}
For $p \geq 1$ and $\gamma\geq 1$, the parameterized promise problem $\gamma$-approximate Closest Vector Problem ($\gamma$-$\GapCVP_p$) is defined as follows.
Instances are tuples $(\mathbf{B}, \vec{t}, d)$, where $\mathbf{B} \in \mathbb{Z}^{m\times n}$ is a lattice basis of $\mathcal{L}(\mathbf{B})$, $\vec{t} \in \mathbb{Z}^m $ is a target vector, and $d \in \mathbb{Z}_{>0}$ is a distance parameter such that
\begin{enumerate}
    \item $(\mathbf{B}, \vec{t}, d)$ is a YES instance if $\operatorname{dist}_p(\mathcal{L}(\mathbf{B}), \vec{t}) \leq d$;
    \item $(\mathbf{B}, \vec{t}, d)$ is a NO instance if $\operatorname{dist}_p(\mathcal{L}(\mathbf{B}), w\vec{t}) > \gamma d$ for all $w\in\mathbb{Z}\backslash\{0\}$.
\end{enumerate}
The parameter of interest is $d$.
\end{definition}

The following theorem states the $\W[1]$-hardness of CVP under deterministic reductions.
\begin{theorem}[{\cite[Theorem 7.2]{journals/BhattacharyyaBE+21}} and {\cite[Theorem 2.11]{journals/sicomp/BennettCGR24}}]
\label{thm:CVP-is-hard}
For any real numbers $\gamma \geq 1$ and $p \geq 1$, $\gamma$-$\GapCVP_p$ is $\W[1]$-hard under deterministic $\FPT$ many-one reductions.
\end{theorem}

For a matrix $\mathbf{G}\in\F_q^{m\times n}$, the linear code generated by the columns of $\mathbf{G}$ is $\mathcal{C}=\mathcal{C}(\mathbf{G}):=\{\mathbf{G}\vec{x}:\vec{x}\in\F_q^n\}\subseteq\F_q^m$.
A parity-check matrix for $\mathcal C$ is a matrix $\mathbf H$ with $\mathcal C=\ker\mathbf H$.
The tensor product of two codes generated by $\mathbf G$ and $\mathbf A$ is the code generated by $\mathbf G\otimes\mathbf A$.
The minimum distance of a nonzero linear code $\mathcal{C}\subseteq\F_q^m$ is $\lambda(\mathcal{C}):=\min_{\vec{c}\in\mathcal{C}\setminus\{\mathbf{0}\}}\|\vec{c}\|_0$.
We use the convention $\lambda(\{\mathbf{0}\}) := +\infty$.
For a target vector $\vec{t}\in\F_q^m$, we write $\operatorname{dist}(\mathcal{C},\vec{t}):=\min_{\vec{c}\in\mathcal{C}}\|\vec{c}-\vec{t}\|_0$.

The Minimum Distance Problem (MDP) is defined as follows.
MDP can be viewed as the coding-theoretic counterpart of SVP.
\begin{definition}[$\gamma$-$\GapMDP_q$, \cite{journals/DumerMS03}]
\label{def:mdp}
For $\gamma \geq 1$ and a prime power $q \geq 2$, the parameterized promise problem $\gamma$-approximate Minimum Distance Problem over $\F_q$ ($\gamma$-$\GapMDP_q$) is defined as follows.
Instances are pairs $(\mathbf{G}, k)$, where $\mathbf{G} \in \F^{m\times n}_q$ is a generator matrix of $\mathcal{C}(\mathbf{G})$, and $k \in \mathbb{Z}_{>0}$ is a distance parameter such that
\begin{enumerate}
    \item $(\mathbf{G}, k)$ is a YES instance if $\lambda(\mathcal{C}(\mathbf{G}))\leq k$;
    \item $(\mathbf{G}, k)$ is a NO instance if $\lambda(\mathcal{C}(\mathbf{G}))> \gamma k$.
\end{enumerate}
The parameter of interest is $k$.
\end{definition}

Similarly to CVP, we next define the Nearest Codeword Problem (NCP), the inhomogeneous version of MDP.
\begin{definition}[$\gamma$-$\GapNCP_q$, \cite{journals/DumerMS03}]
\label{def:ncp}
For $\gamma\geq 1$ and a prime power $q \geq 2$, the parameterized promise problem $\gamma$-approximate Nearest Codeword Problem over $\F_q$ ($\gamma$-$\GapNCP_q$) is defined as follows.
Instances are tuples $(\mathbf{G}, \vec{t}, k)$, where $\mathbf{G} \in \F^{m\times n}_q$ is a generator matrix of $\mathcal{C}(\mathbf{G})$, $\vec{t}\in \F_q^m $ is a target vector, and $k \in \mathbb{Z}_{>0}$ is a distance parameter such that
\begin{enumerate}
    \item $(\mathbf{G}, \vec{t}, k)$ is a YES instance if $\operatorname{dist}(\mathcal{C}(\mathbf{G}), \vec{t}) \leq k$;
    \item $(\mathbf{G}, \vec{t}, k)$ is a NO instance if $\operatorname{dist}(\mathcal{C}(\mathbf{G}), \vec{t}) > \gamma k$.
\end{enumerate}
The parameter of interest is $k$.
\end{definition}

The $\W[1]$-hardness of NCP under deterministic reductions is also known.
\begin{theorem}[{\cite[Theorem 5.1, adapted]{journals/BhattacharyyaBE+21} and \cite[Theorem 2.7]{journals/sicomp/BennettCGR24}}]
\label{thm:NCP-is-hard}
For any real number $\gamma \geq 1$ and any prime power $q \geq 2$, $\gamma$-$\GapNCP_q$ is $\W[1]$-hard under deterministic $\FPT$ many-one reductions.
\end{theorem}

\subsection{Algorithms for finite fields, primes, and lattice bases}
Our reductions require finite fields of suitable sizes and bases for lattices defined by linear congruences. We give deterministic algorithms for these tasks and record the bounds needed to ensure FPT running time.

We represent extension fields using irreducible polynomials.
The following standard result guarantees that such a polynomial exists for every positive degree.
\begin{lemma}[{\cite{LidlNiederreiter1997}}]
\label{lem:irreducible-polynomial-existence}
For every prime power $q$ and every integer $s\ge1$, there exists a monic irreducible polynomial of degree $s$ over $\F_q$.
\end{lemma}

For our purposes, it is enough to construct the field in time polynomial in its number of elements.
The following lemma gives such a construction and shows when the required field computations take FPT time.
\begin{lemma}
\label{lem:explicit-extension-fields}
Fix a prime power $q$.
There exists a deterministic algorithm that, given $s\in\Z_{>0}$, constructs a representation of $\F_{q^s}$ in time $q^{O(s)}$.
In this representation, field arithmetic takes $s^{O(1)}$ time, and all field elements can be enumerated in time $q^{O(s)}$.

Moreover, let $f:\Z_{>0}\to\Z_{>0}$ be computable and let $c\in\Z_{>0}$ be fixed.
If $k,m,s\in\Z_{>0}$ satisfy $q^s\le f(k)m^c$, then constructing the field, enumerating its elements, and performing a number of field operations polynomial in $q^s$ and $s$ take $f_1(k)m^{O(1)}$ time for a computable function $f_1$.
The function $f_1$ and the constant in the exponent depend only on $q$, $f$, $c$, and the fixed polynomial bound on the number of field operations.
\end{lemma}

\begin{proof}
We represent $\F_{q^s}$ as $\F_q[X]/(g)$ for a monic irreducible polynomial $g\in\F_q[X]$ of degree $s$.
Such a polynomial exists by \Cref{lem:irreducible-polynomial-existence}.
Enumerate the monic polynomials of degree $s$ over $\F_q$ in a fixed order.
For each candidate, test whether it is divisible by a monic polynomial of degree between $1$ and $\lfloor s/2\rfloor$, by enumerating all such polynomials and computing the remainders.
A polynomial of degree $s$ is reducible if and only if it has a divisor in this range, so the first candidate that passes these tests is irreducible.
There are $q^s$ candidates and at most $q^s$ possible divisors for each candidate, and each division takes $s^{O(1)}$ time.
Thus the search takes $q^{O(s)}$ time.

Inverses are computed by the extended Euclidean algorithm.
Field arithmetic takes $s^{O(1)}$ time because $q$ is fixed.
Enumerating the $q^s$ coefficient vectors of length $s$ enumerates all field elements in time $q^{O(s)}$.

Any fixed polynomial in $q^s$ and $s$, together with the preceding search time, is at most $(q^s)^{O(1)}$ because $s\le q^s$.
Under $q^s\le f(k)m^c$, this is $f_1(k)m^{O(1)}$ for a computable function $f_1$ with the stated dependence.
\end{proof}

We next explain how to find a prime larger than a given bound.
Bertrand's postulate guarantees that such a prime exists below twice that bound, and deterministic primality testing allows us to find one by checking the candidates in order.
\begin{lemma}[Bertrand's postulate]
\label{lem:bertrand}
For every integer $T\ge2$, there is a prime $Q$ with $T<Q<2T$.
\end{lemma}

\begin{theorem}[{\cite{AgrawalKayalSaxena2004}}]
\label{thm:primality-testing}
There is a deterministic algorithm that, given an integer $N\ge2$, decides whether $N$ is prime in time $(\log N)^{O(1)}$.
\end{theorem}

We can therefore find a prime by testing the integers in the interval in increasing order. The next lemma gives the running time and the condition under which it is FPT.

\begin{lemma}
\label{lem:prime-selection}
There exists a deterministic algorithm that, given an integer $T\ge2$, finds a prime $Q$ with $T<Q<2T$ in time $T^{O(1)}$.
Moreover, let $f:\Z_{>0}\to\Z_{>0}$ be computable and let $c\in\Z_{>0}$ be fixed.
If $d,m\in\Z_{>0}$ satisfy $T\le f(d)m^c$, then the algorithm runs in time $f_1(d)m^{O(1)}$ for a computable function $f_1$ depending only on $f$ and $c$.
\end{lemma}

\begin{proof}
We find the required prime by testing the integers in the specified interval in increasing order.
By \Cref{lem:bertrand}, the interval contains a prime.
For each candidate, apply \Cref{thm:primality-testing}, which takes $(\log T)^{O(1)}$ time.
There are fewer than $T$ candidates, so the search takes $T^{O(1)}$ time.
Substituting $T\le f(d)m^c$ gives the stated bound $f_1(d)m^{O(1)}$.
\end{proof}

For prime $q$, an integer lift of a vector or matrix over $\F_q$ is an integer vector or matrix with the given entries modulo $q$.
Reduction modulo $q$ is applied entrywise.
Finally, we explain how to compute an integer basis for a lattice defined by linear congruences modulo a prime.

\begin{lemma}\label{lem:congruence-kernel-basis}
Let $q$ be prime and let $\mathbf M\in\F_q^{r\times n}$.
A basis of the lattice
\[
 \{\vec v\in\Z^n:\mathbf M(\vec v\bmod q)=\vec0\}
\]
can be computed deterministically in time polynomial in $r,n,\log q$.
\end{lemma}
\begin{proof}
Gaussian elimination, followed by a permutation of columns and deletion of zero rows, puts the matrix in the form $[\mathbf I_h\ \mathbf C]$, where $h$ is its rank.
In that coordinate order a basis is
\[
 \begin{pmatrix}
 q\mathbf I_h&-\widehat{\mathbf C}\\
 \mathbf0&\mathbf I_{n-h}
 \end{pmatrix},
\]
where $\widehat{\mathbf C}$ is the integer lift of $\mathbf C$ with entries in $\{0,\ldots,q-1\}$.
Indeed, the congruence says exactly that $\vec v_1=-\widehat{\mathbf C}\vec v_2+q\vec z$ for an integer vector $\vec z$.
The displayed columns are independent, and undoing the coordinate permutation on every basis vector restores the original coordinate order.
The rank-zero case gives $\mathbf I_n$.
All matrix entries have $O(\log q)$ bits.
\end{proof}

\subsection{Rational points near a nonsingular solution}\label{sec:curve-lemma}

This subsection proves \Cref{lem:keep-all-but-one}, our substitute for the implicit function theorem over finite fields.
We know a common zero $\xi$ of polynomials $f_1,\ldots,f_n$ over a finite field at which their Jacobian matrix is invertible, and we need a common zero of $f_1,\ldots,f_{n-1}$ at which $f_n$ does not vanish.
Over the real numbers, the implicit function theorem would give a curve through $\xi$ on which $f_1,\ldots,f_{n-1}$ stay zero and $f_n$ is nonzero near $\xi$ except at $\xi$.
Over a finite field, we find such a curve algebraically, show that it is defined over the finite field, and count its points.
The subsection assumes no background in algebraic geometry:
we define every notion that we use.
Besides basic linear algebra and finite fields, we use without proof
only \Cref{fact:components}, \Cref{fact:degree},
\Cref{fact:jacobian}, \Cref{thm:curve-points},
and the Frobenius descent criterion
\cite[Example~1.4.2, p.~31]{GM16},
stated in \Cref{lem:frobenius}\ref{item:frob-descent}.

\paragraph{Algebraic sets and curves.}
Throughout this subsection, $K=\F_Q$, and $\overline K$ is an algebraic closure of $K$; every $z\in\overline K$ lies in the finite field $K(z)$.
An \emph{algebraic set} is the set $V(h_1,\ldots,h_s)$ of common zeros in $\overline K^n$ of polynomials $h_1,\ldots,h_s\in\overline K[X_1,\ldots,X_n]$.
Finite sets, finite unions, and finite intersections of algebraic sets are algebraic sets; for instance, $V(g)\cup V(h)=V(gh)$.
An algebraic set $W$ is \emph{defined over $K$} if it equals $V(h_1,\ldots,h_s)$ for some $h_1,\ldots,h_s\in K[X_1,\ldots,X_n]$, and $W(K)=W\cap K^n$ is its set of \emph{$K$-rational points}.
An algebraic set is \emph{irreducible} if it is nonempty and is not the union of two proper algebraic subsets.

\begin{fact}[{see \cite[Section~4.1]{GuoVolkJalanZuckerman2023}}]
\label{fact:components}
Every nonempty algebraic set $V$ is the union of finitely many irreducible algebraic subsets, none of which contains another.
These subsets are uniquely determined by $V$, and they are called the \emph{irreducible components} of $V$.
\end{fact}

Every irreducible algebraic subset $Z$ of $V$ lies in some component $C_i$, since $Z$ is irreducible and is the union of the algebraic sets $Z\cap C_i$.
Hence the components of $V$ are exactly its maximal irreducible algebraic subsets.
An irreducible algebraic set defined over $K$ is called \emph{absolutely irreducible over $K$}.
A \emph{curve} in an algebraic closure is an infinite algebraic set all of whose proper algebraic subsets are finite;\footnote{This is equivalent to the usual definition of a curve as an irreducible algebraic set of dimension one.}
a curve is irreducible, since the union of two finite sets is finite.
A line is an example of a curve.

\paragraph{Degree.}
For a curve $C$, the \emph{degree} is the largest number of points in which $C$ meets an affine hyperplane that does not contain $C$.
We need only the following properties of the degree of a general algebraic set.

\begin{fact}[{see \cite[Section~2]{CafureMatera2006} and \cite[Section~4.1 and Lemma~4.2]{GuoVolkJalanZuckerman2023}}]
\label{fact:degree}
Every algebraic set $V\subseteq\overline K^n$ has a \emph{degree} $\deg V\in\Z_{\ge0}$, which for a curve is the number defined above, with the following properties.
\begin{enumerate}[label=\textup{(\roman*)}]
\item\label{item:degree-finite} The degree of a finite set is its number of points.
\item\label{item:degree-sum} The degree of $V$ is the sum of the degrees of its irreducible components.
\item\label{item:degree-hypersurface} For every nonconstant polynomial $h$, we have $\deg V(h)\le\deg h$.
\item\label{item:degree-bezout} (B\'ezout's inequality) For all algebraic sets $V$ and $W$, we have $\deg(V\cap W)\le\deg V\cdot\deg W$.
\end{enumerate}
\end{fact}

\paragraph{The Frobenius map.}
The Frobenius map $\sigma(z)=z^Q$ is an automorphism of $\overline K$ (it is injective and maps each finite field $K(z)$ into itself), and its fixed points are exactly the $Q$ roots of $Z^Q-Z$, namely the elements of $K$.
We apply $\sigma$ to points coordinatewise, and we write $h^\sigma$ for the polynomial obtained by applying $\sigma$ to the coefficients of $h$.
The following lemma recognizes algebraic sets defined over $K$.

\begin{lemma}\label{lem:frobenius}
\begin{enumerate}[label=\textup{(\roman*)}]
\item\label{item:frob-components} We have $\sigma(V(h_1,\ldots,h_s))=V(h_1^\sigma,\ldots,h_s^\sigma)$, and $\sigma$ maps the irreducible components of every algebraic set $W$ onto those of $\sigma(W)$.
If $W$ is defined over $K$, then $\sigma(W)=W$.
\item\label{item:frob-descent} Conversely, every algebraic set $W$ with $\sigma(W)=W$ is defined over $K$.
\end{enumerate}
\end{lemma}

\begin{proof}
\ref{item:frob-components}
Since $\sigma$ is a ring automorphism, we have $\sigma(h(x))=h^\sigma(\sigma(x))$, so $h_i(x)=0$ if and only if $h_i^\sigma(\sigma(x))=0$; this gives the identity.
Hence $\sigma$ and $\sigma^{-1}$ map algebraic sets to algebraic sets, and $\sigma$ is a bijection between the algebraic subsets of $W$ and those of $\sigma(W)$ that preserves unions and inclusions.
It therefore maps maximal irreducible algebraic subsets to maximal irreducible algebraic subsets.
If $h_i\in K[X_1,\ldots,X_n]$, then $h_i^\sigma=h_i$, and hence $\sigma(W)=W$.

\ref{item:frob-descent}
Since $W$ is an  algebraic subset of $\overline K^n$ and $\sigma(W)=W$, we apply \cite[Example~1.4.2, p.~31]{GM16} with $k\gets\overline K$, $q\gets Q$, and $X\gets W$ to conclude that $W$ is defined over $K$.
\end{proof}

\paragraph{The Jacobian criterion and the Weil bound.}
Over the real numbers, if polynomials $h_1,\ldots,h_r$ have a common zero $b$ at which their Jacobian matrix has rank $r$, the implicit function theorem and the rank theorem (cf. \cite[Theorem 9.32]{Rudin1976}) show that near $b$ their common zero set is a single smooth piece with $n-r$ degrees of freedom.
The Jacobian criterion is the algebraic counterpart of this picture; we need the cases of one and zero degrees of freedom.

\begin{fact}[Jacobian criterion]
\label{fact:jacobian}
Let $h_1,\ldots,h_r\in\overline K[X_1,\ldots,X_n]$, and let $b\in\overline K^n$ be a common zero of them at which their Jacobian matrix has rank $r$.
\begin{enumerate}[label=\textup{(\alph*)}]
\item\label{item:jacobian-curve} If $r=n-1$, then $b$ lies on exactly one irreducible component of $V(h_1,\ldots,h_r)$, and this component is a curve.
\item\label{item:jacobian-point} If $r=n$, then $b$ lies on exactly one irreducible component of $V(h_1,\ldots,h_r)$, namely $\{b\}$.
\end{enumerate}
\end{fact}

\Cref{fact:jacobian} is a standard result in algebraic geometry.\footnote{%
In the language of algebraic geometry: after reordering the variables, the $r\times r$ minor of the Jacobian matrix formed by the first $r$ columns is nonzero at $b$.
Inverting this minor makes $\overline K[X_1,\ldots,X_n]/(h_1,\ldots,h_r)$ a standard smooth algebra \cite[Tag~00T6]{StacksProject}, which is smooth of relative dimension $n-r$ \cite[Tag~00T7]{StacksProject}.
Hence its local ring at $b$ is regular \cite[Tag~056S]{StacksProject} and therefore a domain \cite[Tag~00NP]{StacksProject}, so $b$ lies on exactly one irreducible component \cite[Tag~00ET]{StacksProject}, which has dimension $n-r$.}
The Weil bound below says that a curve defined over $K$ has about as many $K$-rational points as a line defined over $K$, which has exactly $Q$.

\begin{theorem}[Weil bound; {\cite[Theorem~7.1]{CafureMatera2006}}, case of dimension one; see also {\cite[Theorem~4.6]{GuoVolkJalanZuckerman2023}}]
\label{thm:curve-points}
Let $C\subseteq\overline K^n$ be a curve that is defined over $K=\F_Q$, and hence absolutely irreducible over $K$, and let $\delta$ be its degree in the sense of \Cref{fact:degree}.
If $Q>4\delta^2$, then
\[
 \bigl|\,|C(K)|-Q\,\bigr|\le(\delta-1)(\delta-2)\sqrt Q+5\delta^{13/3}.
\]
\end{theorem}

We now prove the lemma.

\begin{lemma}
\label{lem:keep-all-but-one}
Let $K=\F_Q$, let $n\ge2$, and let $f_1,\ldots,f_n\in K[X_1,\ldots,X_n]$ be nonconstant polynomials with $\Delta=\prod_{i=1}^{n-1}\deg f_i$.
Suppose that $\xi\in K^n$ satisfies
\[
 f_1(\xi)=\cdots=f_n(\xi)=0
\]
and that the Jacobian matrix of $f_1,\ldots,f_n$ at $\xi$ is invertible.
If $Q>9\Delta^5$ and $Q>9\Delta\deg f_n$, then there exists $x\in K^n$ such that
\[
 f_1(x)=\cdots=f_{n-1}(x)=0,\qquad f_n(x)\ne0.
\]
\end{lemma}
\begin{proof}
Let $\mathcal V=V(f_1,\ldots,f_{n-1})$. 
We will find a curve $C\subseteq\mathcal V$ defined over $K$
and show that $C$ has more $K$-rational points than zeros of $f_n$.

\emph{Step~1: a curve through $\xi$.}
The first $n-1$ rows of the invertible Jacobian matrix at $\xi$ are linearly independent, and they form the Jacobian matrix of $f_1,\ldots,f_{n-1}$.
By \Cref{fact:jacobian}\ref{item:jacobian-curve}, the point $\xi$ lies on exactly one irreducible component $C$ of $\mathcal V$, and $C$ is a curve.

\emph{Step~2: the curve $C$ is defined over $K$.}
Since $\mathcal V$ is defined over $K$, \Cref{lem:frobenius}\ref{item:frob-components} shows that $\sigma(C)$ is an irreducible component of $\sigma(\mathcal V)$, which is exactly $\mathcal V$.
It contains $\sigma(\xi)=\xi$, since $\xi\in K^n$, so $\sigma(C)=C$ because the point $\xi$ lies on exactly one irreducible component.
By \Cref{lem:frobenius}\ref{item:frob-descent}, the curve $C$ is defined over $K$.\footnote{Wan~\cite[Proposition~3]{Wan1990} uses this argument for the zero set of a single polynomial.}

\emph{Step~3: the polynomial $f_n$ has few zeros on $C$.}
By \Cref{fact:jacobian}\ref{item:jacobian-point}, applied to all $n$ polynomials, the only irreducible component of $V(f_1,\ldots,f_n)$ containing $\xi$ is $\{\xi\}$.
If $f_n$ vanished on all of $C$, then $C$ would be an irreducible algebraic subset of $V(f_1,\ldots,f_n)$ containing $\xi$, so $C\subseteq\{\xi\}$, which is impossible because $C$ is infinite.
Hence $C\cap V(f_n)$ is a proper algebraic subset of $C$, and thus finite.
Let $\delta=\deg C$.
By \Cref{fact:degree}\ref{item:degree-finite}, \ref{item:degree-bezout}, and~\ref{item:degree-hypersurface}, the set $C\cap V(f_n)$ has at most $\deg C\cdot\deg V(f_n)\le\delta\deg f_n$ points.
Similarly, by~\ref{item:degree-bezout} and~\ref{item:degree-hypersurface}, the set $\mathcal V=V(f_1)\cap\cdots\cap V(f_{n-1})$ has degree at most $\Delta$, and by~\ref{item:degree-sum}, so does its component $C$; that is, $\delta\le\Delta$.
Since $\xi\in C\cap V(f_n)$, we also have $1\le\delta\deg f_n$, so $\delta\ge1$.

\emph{Step~4: the curve $C$ has many points in $K^n$.}
Since $1\le\delta\le\Delta$, we have $Q>9\Delta^5\ge4\delta^2$, $(\delta-1)(\delta-2)\le\Delta^2$, and $\delta^{13/3}\le\Delta^5$.
By Step~2, \Cref{thm:curve-points} applies to $C$ and gives
\[
 |C(K)|\ge Q-\Delta^2\sqrt Q-5\Delta^5>Q-\frac Q3-\frac{5Q}9=\frac Q9>\Delta\deg f_n,
\]
where we used $\sqrt Q>3\Delta^2$ and $5\Delta^5<5Q/9$, which follow from $Q>9\Delta^5$, and the assumption $Q>9\Delta\deg f_n$.
By Step~3, some $x\in C(K)$ is not a zero of $f_n$, and $f_1(x)=\cdots=f_{n-1}(x)=0$ because $C\subseteq\mathcal V$.
\end{proof}

\section{A Deterministic Reduction from CVP to SVP}\label{sec:svp}

We prove \Cref{thm:SVP-is-W[1]hard} by constructing a deterministic reduction from CVP to SVP.
We observe that the reduction of Hirahara and Ogitsuka~\cite{conf/random/HiraharaO26} becomes deterministic if we can deterministically construct a locally dense lattice, as defined below, in $\FPT$ time.
Our construction uses the Reed--Solomon lattice of Bennett and Peikert~\cite{conf/random/BennettP23}.

\begin{definition}[Locally dense lattice]\label{def:locally-dense-lattice}
Let $p\ge1$, let $m,d,D,W$ be positive integers, and let $q$ be prime.
Let $\mathbf{H}\in\F_q^{r\times N}$, let $\vec{s}\in\Z^N$, and let $S$ be an ordered list of $m$ distinct coordinates.
Put
\[
 L=\{\vec{v}\in\Z^N:\mathbf{H}(\vec{v}\bmod q)=\vec{0}\}.
\]
The tuple $(q,\mathbf{H},\vec{s},S)$ is a locally dense lattice with parameters $(p,m,d,D,W)$ if
\begin{enumerate}[label=\textup{(\roman*)}]
\item\label{item:ldl-minimum-distance} $\lambda_1^{(p)}(L)>D$;
\item\label{item:ldl-center} $\vec{s}|_S=\vec{0}$;
\item\label{item:ldl-local-density} for every $\vec{u}\in\Z^m$ with $\|\vec{u}\|_p\le d$, there exists $\vec{w}\in L-\vec{s}$ such that
\[
 \vec{w}|_S=\vec{u},\qquad \|\vec{w}\|_p^p\le W.
\]
\end{enumerate}
Here $W$ bounds the $p$th power of the norm, while $d$ and $D$ are norm thresholds.
The equality $\vec{w}|_S=\vec{u}$ in condition~\ref{item:ldl-local-density} is over the integers.
Since $L$ contains $q$ times every unit vector, condition~\ref{item:ldl-minimum-distance} implies $q>D$.
\end{definition}

\subsection{A Non-Uniform Reduction Using a Locally Dense Lattice}

The following reduction is essentially based on the observation that the reduction of Hirahara and Ogitsuka gives a non-uniform reduction when a locally dense lattice is provided as advice.
For fixed constants of the reduction, the same locally dense lattice can be used for all instances with the same $m$ and $d$.

\begin{theorem}[{\cite[proof of Theorem~3.9 in the full version, adapted]{conf/random/HiraharaO26}}]
\label{thm:CVP-to-SVP}
Fix $p\ge1$, $\gamma'\ge1$, and integers $\gamma\ge2$, $c\ge1$ with $\gamma'c<\gamma$.
There is a deterministic algorithm that, given an instance $(\mathbf{B},\vec{t},d)$ of $\gamma$-$\GapCVP_p$ with $\mathbf{B}\in\Z^{m\times n}$ and a locally dense lattice $(q,\mathbf{H},\vec{s},S)$ with parameters $(p,m,d,\gamma d,W)$ satisfying
\begin{equation}\label{eq:svp-norm-budget}
 d^p+W+1\le(cd)^p,
\end{equation}
outputs an instance of $\gamma'$-$\GapSVP_p$ with parameter $d'=cd$, preserving both YES and NO instances.
Its running time is polynomial in the total input length, including the representation of the supplied locally dense lattice.
\end{theorem}

\begin{proof}
We first construct the output instance.
Set $D=\gamma d$ and $\beta=D+1$.
Use \Cref{lem:congruence-kernel-basis} to compute an integer basis $\mathbf{A}\in\Z^{N\times N}$ of the lattice $L$ represented by $q$ and $\mathbf{H}$.
Let $\mathbf{A}_S$ consist of the rows of $\mathbf{A}$ indexed by $S$, in the specified order.
Output $(\mathbf{B}',d')$, where
\begin{equation}\label{eq:svp-output-basis}
 \mathbf{B}'=
 \begin{pmatrix}
  \mathbf{B} & \mathbf{0} & -\vec{t}\\
  \mathbf{0} & \mathbf{A} & -\vec{s}\\
  \mathbf{0} & \mathbf{0} & 1\\
  \beta\mathbf{B} & -\beta\mathbf{A}_S & -\beta\vec{t}
 \end{pmatrix},
 \qquad d'=cd.
\end{equation}
Compared with~\cite{conf/random/HiraharaO26}, we omit the columns $\beta q\mathbf{I}_m$, since the local density condition in \Cref{def:locally-dense-lattice} holds as an equality over the integers.

We show that the columns of $\mathbf{B}'$ are linearly independent, so that $\mathbf{B}'$ is a lattice basis.
Consider a linear combination of its columns that is zero.
The third block forces the coefficient of the column containing $-\vec{t}$ and $-\vec{s}$ to be zero.
Since $\mathbf{B}$ and $\mathbf{A}$ have linearly independent columns, the first two blocks then force the coefficients of their columns to be zero.

Next, we show completeness.
Suppose that $\|\mathbf{B}\vec{x}-\vec{t}\|_p\le d$ for some $\vec{x}\in\Z^n$.
Since $\lat(\mathbf{A})=L$, the local density condition in \Cref{def:locally-dense-lattice} gives $\vec{y}\in\Z^N$ such that
\[
 \|\mathbf{A}\vec{y}-\vec{s}\|_p^p\le W,\qquad
 (\mathbf{A}\vec{y}-\vec{s})|_S=\mathbf{B}\vec{x}-\vec{t}.
\]
Since $\vec{s}|_S=\vec{0}$, the last equality gives $\mathbf{A}_S\vec{y}=\mathbf{B}\vec{x}-\vec{t}$.
Therefore,
\[
 \mathbf{B}'
 \begin{pmatrix}
  \vec{x}\\
  \vec{y}\\
  1
 \end{pmatrix}
 =
 \begin{pmatrix}
  \mathbf{B}\vec{x}-\vec{t}\\
  \mathbf{A}\vec{y}-\vec{s}\\
  1\\
  \vec{0}
 \end{pmatrix}.
\]
This vector is nonzero because its third block is $1$.
By \Cref{eq:svp-norm-budget}, the $p$th power of its norm satisfies
\[
 \|\mathbf{B}\vec{x}-\vec{t}\|_p^p
 +\|\mathbf{A}\vec{y}-\vec{s}\|_p^p+1
 \le d^p+W+1\le(cd)^p.
\]
Thus the output is a YES instance with parameter $d'=cd$.

We now show soundness.
Suppose that $(\mathbf{B},\vec{t},d)$ is a NO instance.
By \Cref{def:cvp}, this means that
\[
 \operatorname{dist}_p(\lat(\mathbf{B}),\ell\vec{t})>D
 \qquad(\ell\in\Z\setminus\{0\}).
\]
Assume for contradiction that $\lat(\mathbf{B}')$ contains a nonzero vector of norm at most $D$.
Using $\vec{s}|_S=\vec{0}$, we can write this vector as
\[
 \begin{pmatrix}
  \mathbf{B}\vec{x}-\ell\vec{t}\\
  \mathbf{A}\vec{y}-\ell\vec{s}\\
  \ell\\
  \beta\bigl(\mathbf{B}\vec{x}-\ell\vec{t}
       -(\mathbf{A}\vec{y}-\ell\vec{s})|_S\bigr)
 \end{pmatrix}
\]
for some $\vec{x}\in\Z^n$, $\vec{y}\in\Z^N$, and $\ell\in\Z$.
If $\ell\ne0$, the first block has norm
\[
 \|\mathbf{B}\vec{x}-\ell\vec{t}\|_p
 \ge\operatorname{dist}_p(\lat(\mathbf{B}),\ell\vec{t})>D.
\]
This contradicts the assumed norm bound, so $\ell=0$.

The case $\ell=0$ also leads to a contradiction.
The second block is now $\mathbf{A}\vec{y}\in L$ and has norm at most $D$.
Since $\lambda_1^{(p)}(L)>D$, we must have $\mathbf{A}\vec{y}=\vec{0}$.
The last block is therefore $\beta\mathbf{B}\vec{x}$, and every coordinate in it is an integer multiple of $\beta>D$.
The assumed norm bound therefore forces that block to be zero, so $\mathbf{B}\vec{x}=\vec{0}$.
All blocks are now zero, contrary to the choice of a nonzero vector.
Therefore,
\[
 \lambda_1^{(p)}(\lat(\mathbf{B}'))>D=\gamma d>\gamma'cd.
\]

We finally bound the running time.
The matrix $\mathbf{B}'$ has $N+2m+1$ rows and $n+N+1$ columns.
By \Cref{lem:congruence-kernel-basis}, computing $\mathbf{A}$ takes time polynomial in the dimensions of $\mathbf{H}$ and $\log q$.
Forming $\mathbf{B}'$ also takes time polynomial in the total input length, and the bit lengths of all its entries are polynomial in that length.
Thus the reduction has the claimed running time.
\end{proof}

\subsection{A Deterministic FPT Construction of Locally Dense Lattices}

In this section, we give a deterministic $\FPT$ algorithm that constructs a locally dense lattice as defined in \Cref{def:locally-dense-lattice}.
Our construction uses a Reed--Solomon lattice.
We use the following lower bound on its minimum norm to verify condition~\ref{item:ldl-minimum-distance} of \Cref{def:locally-dense-lattice}.
The proof of the minimum distance in \cite[Theorem~3.1 in the full version]{conf/random/BennettP23} applies without change to extension fields under the characteristic assumption below.
We include the proof for completeness.
\begin{lemma}[{\cite[Theorem~3.1 in the full version, adapted]{conf/random/BennettP23}}]
\label{lem:svp-rs-lattice}
Let $k\ge2$ be an integer, let $q>2k$ be prime, and let $Q=q^s$ for an integer $s\ge1$.
The lattice
\begin{equation}\label{eq:svp-rs-lattice}
 L_{Q,k}=\left\{\vec{v}\in\Z^{\F_Q}:
   \sum_{z\in\F_Q}v_z z^j=0\text{ in }\F_Q
   \quad(0\le j<k)\right\}
\end{equation}
satisfies
\[
 \lambda_1^{(p)}(L_{Q,k})\ge(2k)^{1/p},
 \qquad \lambda(L_{Q,k}\bmod q)\ge k+1
\]
for every real $p\ge1$.
The integer coefficients are interpreted in $\F_q\subseteq\F_Q$.
The equation for $j=0$ uses $z^0=1$ for every $z\in\F_Q$.
\end{lemma}

\begin{proof}
We first bound the minimum $\ell_1$ norm.
Suppose that $\vec{0}\ne\vec{v}\in L_{Q,k}$ satisfies $\|\vec{v}\|_1<2k$.
The equation for $j=0$ gives $\sum_z v_z\equiv0\pmod{q}$.
Since $|\sum_z v_z|\le\|\vec{v}\|_1<2k<q$, we have $\sum_z v_z=0$ over the integers.
Form two multisets of elements of $\F_Q$: for each positive entry $v_z$, put $v_z$ copies of $z$ in the first multiset, and for each negative entry $v_z$, put $-v_z$ copies of $z$ in the second.
Their cardinalities are equal, say to $h$, and $2h=\|\vec{v}\|_1<2k$.
The case $h=0$ would give $\vec{v}=\vec{0}$, so $1\le h<k$.

The equations for $j=1,\ldots,h$ in \Cref{eq:svp-rs-lattice} say that these multisets have equal power sums through degree $h$.
Since the characteristic $q$ exceeds $h$, Newton's identities recover their elementary symmetric functions from these power sums.
The monic polynomials having the elements of the two multisets as roots, counted with multiplicity, are therefore equal.
Unique factorization in $\F_Q[X]$ implies that the two multisets are equal.
This is impossible because the positive and negative supports of $\vec{v}$ are disjoint and $h\ge1$.
Thus $\lambda_1^{(1)}(L_{Q,k})\ge2k$.
For every integer vector $\vec{v}$ and every real $p\ge1$, we have $\|\vec{v}\|_p^p\ge\|\vec{v}\|_1$.
This gives $\lambda_1^{(p)}(L_{Q,k})\ge(2k)^{1/p}$.

Next, suppose that a nonzero word of $L_{Q,k}\bmod q$ has support $\{z_1,\ldots,z_h\}$ of size $h\le k$.
The first $h$ equations in \Cref{eq:svp-rs-lattice} have coefficient matrix
\[
 (z_i^j)_{0\le j<h,\,1\le i\le h}.
\]
This Vandermonde matrix is invertible over $\F_Q$ because the $z_i$ are distinct.
All coefficients on the support must therefore be zero, a contradiction.
This proves $\lambda(L_{Q,k}\bmod q)\ge k+1$.
\end{proof}

Taking $s=1$ gives the lattice $L_{q,k}$ over the prime field $\F_q$.

We now choose the shift $\vec{s}$ and the coordinates $S$ so that the Reed--Solomon lattice satisfies the local density condition.
The following theorem gives a deterministic construction for the integer parameters specified below.

\begin{theorem}\label{thm:rs-locally-dense-lattice}
Fix $p\ge1$.
Let $m,d,D,h,k$ be positive integers satisfying
\begin{equation}\label{eq:svp-aux-conditions}
 D\ge2,\qquad k\ge2,\qquad d^p\le h,\qquad D^p<2k,
\end{equation}
and put $W=k-1+2h$.
There is a deterministic algorithm that, on input $(m,d,D,h,k)$, constructs a locally dense lattice $(q,\mathbf{H},\vec{s},S)$ with parameters $(p,m,d,D,W)$.
The algorithm outputs the prime $q$, the matrix $\mathbf{H}$, the vector $\vec{s}$, and the coordinates in $S$.
Its running time and output size are bounded by $f(d+D+h+k)(m+1)^{O(1)}$ for a computable function $f$, with an exponent independent of $d,D,h,k$.
\end{theorem}
\begin{proof}
\setcounter{claim}{0}
We first choose the prime field that indexes the coordinates of the lattice.
Put $L=k-1$ and use \Cref{lem:prime-selection} to choose a prime $q$ with $T<q<2T$, where
\begin{equation}\label{eq:svp-prime-bound}
 \begin{aligned}
 T=1+\max\bigl\{2k,\ D,\ m+L,\ 9(L!)^5,9L!\left(1+(L+h)m+Lh+\binom{L}{2}\right)\bigr\}.
 \end{aligned}
\end{equation}
For $z\in\F_q$, let $\vec{e}_z\in\Z^{\F_q}$ have entry $1$ at $z$ and entry $0$ at every other coordinate.

We next construct the matrix $\mathbf{H}$, the vector $\vec{s}$, and the ordered list of coordinates $S$.
Index the coordinates by $\F_q$ and define $\mathbf{H}=(z^j)_{0\le j<k,\,z\in\F_q}$, with $z^0=1$ for every $z$.
Then $\{\vec{v}\in\Z^{\F_q}:\mathbf{H}(\vec{v}\bmod q)=\vec{0}\}$ is the lattice $L_{q,k}$ in \Cref{eq:svp-rs-lattice}.
Since $q>m+L$, we can choose distinct elements $y_1,\ldots,y_L,x_1,\ldots,x_m\in\F_q$ using a fixed enumeration of $\F_q$.
Set
\[
 \vec{s}=-\sum_{\ell=1}^L\vec{e}_{y_\ell},\qquad S=(x_1,\ldots,x_m).
\]
Thus $\vec{s}|_S=\vec{0}$.
The algorithm outputs $(q,\mathbf{H},\vec{s},S)$.
When considering membership in $S$, we identify $S$ with its underlying set $\{x_1,\ldots,x_m\}$.

We first verify the minimum distance condition in \Cref{def:locally-dense-lattice}.
\begin{claim}\label{clm:rs-distance}
The lattice $L_{q,k}$ satisfies $\lambda_1^{(p)}(L_{q,k})>D$.
\end{claim}
\begin{proof}[Proof of \Cref{clm:rs-distance}]
We apply \Cref{lem:svp-rs-lattice} with $Q=q$ to bound the minimum norm.
Its hypotheses hold because $k\ge2$ and $q>2k$.
Together with $D^p<2k$, it gives $\lambda_1^{(p)}(L_{q,k})\ge(2k)^{1/p}>D$.
\end{proof}

We now construct, for each $\vec{u}\in\Z^m$ with $\|\vec{u}\|_p\le d$, a vector $\vec{w}\in L_{q,k}-\vec{s}$ satisfying
\[
 \vec{w}|_S=\vec{u},\qquad \|\vec{w}\|_p^p\le W,
\]
as required by condition~\ref{item:ldl-local-density} of \Cref{def:locally-dense-lattice}.
If $\vec{u}=\vec{0}$, choose $\vec{w}=-\vec{s}$.
Then $\vec{w}+\vec{s}=\vec{0}\in L_{q,k}$, $\vec{w}|_S=\vec{0}$, and $\|\vec{w}\|_p^p=L\le W$.
For the rest of the proof, fix $\vec{u}\ne\vec{0}$ and put $I=\{i\in[m]:u_i\ne0\}$.
Since the entries of $\vec{u}$ are integers, every nonzero entry satisfies $|u_i|^p\ge1$, so
\[
 1\le|I|\le\sum_{i=1}^m|u_i|^p\le d^p\le h.
\]
For any $z_1,\ldots,z_L,\tau\in\F_q$, define
\begin{equation}\label{eq:svp-completion-vector}
 \vec{w}=\sum_{\ell=1}^L\vec{e}_{z_\ell}
       +\sum_{i\in I}u_i(\vec{e}_{x_i}-\vec{e}_{x_i+\tau})
       \in\Z^{\F_q}.
\end{equation}
The additions in the coordinate indices are taken in $\F_q$, while the coefficients of $\vec{w}$ are integers.

We obtain the required norm bound whenever the coordinate indices in \Cref{eq:svp-completion-vector} are pairwise distinct.
If the elements in the list $z_1,\ldots,z_L,(x_i)_{i\in I},(x_i+\tau)_{i\in I}$ are pairwise distinct, then the nonzero entries of $\vec{w}$ are $L$ copies of $1$ and one copy each of $u_i$ and $-u_i$ for every $i\in I$.
Hence
\[
 \|\vec{w}\|_p^p=L+2\sum_{i\in I}|u_i|^p
 \le k-1+2h=W.
\]

We derive the equations equivalent to $\vec{w}\in L_{q,k}-\vec{s}$.
By \Cref{eq:svp-rs-lattice}, the condition for $j=0$ is
\[
 \sum_{z\in\F_q}(w_z+s_z)=0\pmod{q}.
\]
This condition holds for every choice of $z_1,\ldots,z_L,\tau\in\F_q$, since
\[
 \sum_{z\in\F_q}(w_z+s_z)
 =\sum_{\ell=1}^L(1-1)+\sum_{i\in I}u_i(1-1)=0
\]
over the integers.
By \Cref{eq:svp-rs-lattice}, the remaining conditions are
\[
 \sum_{z\in\F_q}(w_z+s_z)z^j=0\pmod{q}
 \qquad(1\le j\le L).
\]
We interpret the integer coefficients in $\F_q$ when evaluating these sums.
For each $1\le j\le L$ and $\ell\in[L]$, we have
\[
 \sum_{z\in\F_q}(\vec{e}_{z_\ell})_z z^j=z_\ell^j,
\]
since $\vec{e}_{z_\ell}$ has entry $1$ at $z_\ell$ and entry $0$ at every other coordinate.
Applying the same calculation to each term of $\vec{w}+\vec{s}$ gives
\[
 \begin{aligned}
 \sum_{z\in\F_q}(w_z+s_z)z^j
 &=\sum_{\ell=1}^L(z_\ell^j-y_\ell^j)
   +\sum_{i\in I}u_i\bigl(x_i^j-(x_i+\tau)^j\bigr)\\
 &=\sum_{\ell=1}^L(z_\ell^j-y_\ell^j)
   -\sum_{i\in I}u_i\bigl((x_i+\tau)^j-x_i^j\bigr).
 \end{aligned}
\]
Let $Z_1,\ldots,Z_L,Y$ be variables over $\F_q$.
Therefore, $\vec{w}+\vec{s}\in L_{q,k}$ if and only if $(z_1,\ldots,z_L,\tau)$ satisfies
\begin{equation}\label{eq:svp-completion-equations}
 \sum_{\ell=1}^L(Z_\ell^j-y_\ell^j)
 -\sum_{i\in I}u_i\bigl((x_i+Y)^j-x_i^j\bigr)=0
 \qquad(1\le j\le L),
\end{equation}
where the coefficients $u_i$ are interpreted in $\F_q$.

To ensure $\vec{w}|_S=\vec{u}$ and $\|\vec{w}\|_p^p\le W$, we introduce a polynomial $g$.
Its nonzero values will ensure that every $z_\ell$ and every $x_i+\tau$ with $i\in I$ lies outside $S$ and that all coordinate indices in \Cref{eq:svp-completion-vector} are pairwise distinct.
Set
\[
 E=\{x_b-x_i:i\in I,\ b\in[m]\}.
\]
Since $I\ne\varnothing$, we have $0\in E$ and $|E|\le hm$.
Define the polynomial $g\in\F_q[Z_1,\ldots,Z_L,Y]$ by
\begin{equation}\label{eq:svp-last-polynomial}
 \begin{aligned}
 g(Z_1,\ldots,Z_L,Y)
 &=Y\prod_{e\in E\setminus\{0\}}(Y-e)
       \prod_{\ell=1}^L\prod_{b=1}^m(Z_\ell-x_b)\\
 &\quad\cdot\prod_{1\le\ell<\ell'\le L}(Z_\ell-Z_{\ell'})
       \prod_{\ell=1}^L\prod_{i\in I}(Z_\ell-x_i-Y).
 \end{aligned}
\end{equation}
The empty product is defined to be $1$.

\begin{claim}\label{clm:rs-restriction}
For every $z_1,\ldots,z_L,\tau\in\F_q$ with $g(z_1,\ldots,z_L,\tau)\ne0$, we have
\[
 \begin{gathered}
 z_\ell\notin S\quad(\ell\in[L]),\qquad
 x_i+\tau\notin S\quad(i\in I),\\
 z_\ell\ne z_{\ell'}\quad(1\le\ell<\ell'\le L),\qquad
 z_\ell\ne x_i+\tau\quad(\ell\in[L],\ i\in I).
 \end{gathered}
\]
Consequently, the vector in \Cref{eq:svp-completion-vector} satisfies $\vec{w}|_S=\vec{u}$ and $\|\vec{w}\|_p^p\le W$.
\end{claim}
\begin{proof}[Proof of \Cref{clm:rs-restriction}]
We first show that $z_\ell\notin S$ for every $\ell\in[L]$.
Since $g(z_1,\ldots,z_L,\tau)\ne0$, every factor of $g$ is nonzero at $(z_1,\ldots,z_L,\tau)$.
The factors $Z_\ell-x_b$ therefore give $z_\ell\ne x_b$ for every $\ell\in[L]$ and $b\in[m]$.
Since $S$ consists of $x_1,\ldots,x_m$, this proves $z_\ell\notin S$.

We next show that $x_i+\tau\notin S$ for every $i\in I$.
The factor $Y$ gives $\tau\ne0$, and the factors $Y-e$ give $\tau\ne e$ for every $e\in E\setminus\{0\}$.
Thus $\tau\notin E$.
For every $i\in I$ and $b\in[m]$, the definition of $E$ gives $x_b-x_i\in E$.
Hence $\tau\ne x_b-x_i$, which implies $x_i+\tau\ne x_b$.
Therefore $x_i+\tau\notin S$ for every $i\in I$.

The remaining factors give the other two groups of inequalities.
The factors $Z_\ell-Z_{\ell'}$ give $z_\ell\ne z_{\ell'}$ whenever $\ell<\ell'$.
The factors $Z_\ell-x_i-Y$ give $z_\ell\ne x_i+\tau$ for every $\ell\in[L]$ and $i\in I$.

We now show that $\vec{w}|_S=\vec{u}$ over the integers.
The vectors $\vec{e}_{z_\ell}$ and $\vec{e}_{x_i+\tau}$ have entry $0$ at every coordinate in $S$, as their indices lie outside $S$.
By \Cref{eq:svp-completion-vector}, the entries of $\vec{w}$ on $S$ are therefore the same as those of $\sum_{i\in I}u_i\vec{e}_{x_i}$.
Since $x_1,\ldots,x_m$ are distinct, this gives $w_{x_b}=u_b$ for every $b\in I$.
For $b\in[m]\setminus I$, the sum has entry $0$ at $x_b$, and $u_b=0$ by the definition of $I$.
Thus $w_{x_b}=u_b$ for every $b\in[m]$, which proves $\vec{w}|_S=\vec{u}$.

We finally show that $\|\vec{w}\|_p^p\le W$.
The elements $x_i$ with $i\in I$ are distinct, and adding the same $\tau$ preserves distinctness.
Together with the inequalities proved above, this shows that all coordinate indices in \Cref{eq:svp-completion-vector} are pairwise distinct.
The preceding norm calculation therefore gives $\|\vec{w}\|_p^p\le W$.
\end{proof}

To complete the construction of $\vec{w}$, it is enough to show that there exists a point $(z_1,\ldots,z_L,\tau)\in\F_q^{L+1}$ satisfying \Cref{eq:svp-completion-equations} and $g(z_1,\ldots,z_L,\tau)\ne0$, where $g$ is defined in \Cref{eq:svp-last-polynomial}.
For such a point, the vector $\vec{w}$ defined in \Cref{eq:svp-completion-vector} belongs to $L_{q,k}-\vec{s}$.
By \Cref{clm:rs-restriction}, it also satisfies $\vec{w}|_S=\vec{u}$ and $\|\vec{w}\|_p^p\le W$.
We prove that such a point exists by applying \Cref{lem:keep-all-but-one}.

We first show that $\xi=(y_1,\ldots,y_L,0)$ is a common zero of the left sides of \Cref{eq:svp-completion-equations} and $g$.
At $\xi$, we have $Z_\ell=y_\ell$ for every $\ell\in[L]$ and $Y=0$.
Thus every left side in \Cref{eq:svp-completion-equations} is zero, since both $Z_\ell^j-y_\ell^j$ and $(x_i+Y)^j-x_i^j$ are zero at this point.
Since $g$ has the factor $Y$, we also have $g(\xi)=0$.

We next show that the Jacobian matrix of these $L+1$ polynomials is invertible at $\xi$.
We order the $L$ left sides by increasing $j$, followed by $g$, and use the variable order $(Z_1,\ldots,Z_L,Y)$.
For the left side indexed by $j$, the derivative with respect to $Z_\ell$ is $jZ_\ell^{j-1}$, whose value at $\xi$ is $jy_\ell^{j-1}$.
Thus the upper left block is
\[
 \begin{pmatrix}
  1 & 1 & \cdots & 1\\
  2y_1 & 2y_2 & \cdots & 2y_L\\
  \vdots & \vdots & \ddots & \vdots\\
  Ly_1^{L-1} & Ly_2^{L-1} & \cdots & Ly_L^{L-1}
 \end{pmatrix}.
\]
This is a Vandermonde matrix with its $j$th row multiplied by $j$.
The elements $y_1,\ldots,y_L$ are distinct, and $1,\ldots,L$ are nonzero in $\F_q$ because $q>L$.
Hence this block is invertible.

We use the factor $Y$ in $g$ to compute the last row of the Jacobian matrix.
Each derivative of $g$ with respect to $Z_\ell$ still has the factor $Y$, so
\[
 \frac{\partial g}{\partial Z_\ell}(\xi)=0
 \qquad(\ell\in[L]).
\]
When differentiating $g$ with respect to $Y$, every term except the one obtained by differentiating the first factor $Y$ still contains $Y$ and is zero at $\xi$.
Therefore,
\[
 \begin{aligned}
 \frac{\partial g}{\partial Y}(\xi)
 =\prod_{e\in E\setminus\{0\}}(-e)
       \prod_{\ell=1}^L\prod_{b=1}^m(y_\ell-x_b)\quad\cdot\prod_{1\le\ell<\ell'\le L}(y_\ell-y_{\ell'})
       \prod_{\ell=1}^L\prod_{i\in I}(y_\ell-x_i)\ne0.
 \end{aligned}
\]
Every factor in this expression is nonzero because $e\ne0$ and the elements $y_1,\ldots,y_L,x_1,\ldots,x_m$ are distinct.
The Jacobian matrix thus has an invertible upper left block, a zero lower left block, and a nonzero lower right entry, so it is invertible.

We now check the remaining assumptions of \Cref{lem:keep-all-but-one} concerning the degrees and the size of $\F_q$.
For each $1\le j\le L$, every term on the left side of \Cref{eq:svp-completion-equations} has degree at most $j$, and the coefficient of $Z_1^j$ is $1$.
Its degree is therefore exactly $j$, and the product of these degrees is $L!$.
Since $g$ is a product of linear factors, we have
\[
 \begin{aligned}
 \deg g
 &=1+|E\setminus\{0\}|+Lm+\binom L2+L|I|\\
 &\le1+(L+h)m+Lh+\binom L2.
 \end{aligned}
\]
All $L+1$ polynomials are nonconstant.
Our choice of $q$ in \Cref{eq:svp-prime-bound} gives
\[
 \begin{aligned}
 q&>9(L!)^5,\\
 q&>9L!\left(1+(L+h)m+Lh+\binom{L}{2}\right)
   \ge9L!\deg g.
 \end{aligned}
\]

We can now apply \Cref{lem:keep-all-but-one} to obtain the required point.
We use the $L$ left sides of \Cref{eq:svp-completion-equations}, followed by $g$, as the polynomials in the lemma, with variables $(Z_1,\ldots,Z_L,Y)$.
In this application, $n=L+1=k\ge2$, $Q=q$, $\Delta=L!$, and $f_n=g$.
The lemma gives a point $(z_1,\ldots,z_L,\tau)\in\F_q^{L+1}$ satisfying \Cref{eq:svp-completion-equations} and $g(z_1,\ldots,z_L,\tau)\ne0$.

The point given by the lemma completes the construction of $\vec{w}$ for the fixed nonzero vector $\vec{u}$.
Together with the case $\vec{u}=\vec{0}$ treated above, this proves condition~\ref{item:ldl-local-density} of \Cref{def:locally-dense-lattice}.
\Cref{clm:rs-distance} and $\vec{s}|_S=\vec{0}$ give the other two conditions.

We finally bound the running time and the size of the output.
The algorithm finds $q$, computes the matrix $\mathbf{H}$, and outputs $q$, $\mathbf{H}$, $S$, and $\vec{s}$.
Since $L=k-1$, the bound in \Cref{eq:svp-prime-bound} gives
\[
 T\le f_0(d+D+h+k)(m+1)
\]
for a computable function $f_0$.
By \Cref{lem:prime-selection}, finding a prime $q$ with $T<q<2T$ takes time polynomial in $T$.
The matrix $\mathbf{H}$ has $k$ rows and $q$ columns, and each entry has $O(\log q)$ bits.
Computing this matrix and choosing $\vec{s}$ and $S$ take time polynomial in $q$ and the integer parameters.
The output size satisfies the same bound.
Therefore, after increasing the computable function if necessary, both the running time and the output size are at most $f(d+D+h+k)(m+1)^{O(1)}$, with an exponent independent of $d,D,h,k$.
\end{proof}

\subsection{Finishing the Proof}

We now combine the reduction with the deterministic construction of a locally dense lattice.
To obtain any fixed approximation factor $\gamma'<2^{1/p}$, we choose the parameters so that the constructed lattice satisfies the minimum norm condition and the bound in \Cref{eq:svp-norm-budget}.
The following theorem gives the resulting deterministic $\FPT$ reduction from CVP to SVP.

\begin{theorem}\label{thm:CVP-to-SVP-deterministic}
Fix a computable real $p\ge1$ and a constant $\gamma'\in[1,2^{1/p})$.
There are integers $\gamma\ge2$, $c\ge1$ and a deterministic $\FPT$ many-one reduction from $\gamma$-$\GapCVP_p$ to $\gamma'$-$\GapSVP_p$.
The output parameter is $d'=cd$, where $d$ is the input parameter.
\end{theorem}
\begin{proof}
We first choose fixed integers $\gamma$ and $c$ that give the required approximation factor.
Choose them so that
\begin{equation}\label{eq:svp-gap-constants}
 \gamma\ge2,\qquad c\ge1,\qquad
 \gamma^p/2+8\le c^p,\qquad \gamma'c<\gamma.
\end{equation}
Such integers exist because, as $\gamma$ tends to infinity through the positive integers,
\[
 \frac{\lceil(\gamma^p/2+8)^{1/p}\rceil}{\gamma}
 \longrightarrow 2^{-1/p}<\frac{1}{\gamma'}.
\]
Thus, for all sufficiently large integers $\gamma$, taking $c=\lceil(\gamma^p/2+8)^{1/p}\rceil$ satisfies the conditions in \Cref{eq:svp-gap-constants}.
Fix one such pair $\gamma,c$ for the reduction.

We next compute the integer parameters needed to construct the locally dense lattice.
Given an instance $(\mathbf{B},\vec{t},d)$ of $\gamma$-$\GapCVP_p$ with $\mathbf{B}\in\Z^{m\times n}$, put $D=\gamma d$ and choose integers $h,k$ with
\begin{equation}\label{eq:svp-rounded-parameters}
 d^p\le h<d^p+3/2,\qquad D^p/2<k<D^p/2+3/2.
\end{equation}
To compute these integers, first approximate $d^p$ and $D^p/2$ from above by rational numbers, with a positive error less than $1/2$ in each case.
Taking the ceilings of these rational numbers gives $h$ and $k$ satisfying the displayed inequalities.
Since $p$ is computable and $\gamma$ is fixed, this computation takes time bounded by a computable function of $d$.

We verify that these parameters satisfy the assumptions of \Cref{thm:rs-locally-dense-lattice}.
Since $D=\gamma d\ge2$ and $p\ge1$, we have $D^p/2\ge1$.
Thus $k>D^p/2$ implies $k\ge2$ and $D^p<2k$.
Our choice of $h$ also gives $d^p\le h$.
Apply \Cref{thm:rs-locally-dense-lattice} to construct a locally dense lattice with parameters $(p,m,d,D,W)$, where $W=k-1+2h$.

We now verify the norm bound needed to apply \Cref{thm:CVP-to-SVP}.
By \Cref{eq:svp-rounded-parameters}, we have
\[
 \begin{aligned}
 d^p+W+1
 &=d^p+k+2h\\
 &<D^p/2+3d^p+9/2\\
 &\le(\gamma^p/2+8)d^p\\
 &\le(cd)^p.
 \end{aligned}
\]
Here the second inequality uses $D=\gamma d$ and $d^p\ge1$, and the last inequality follows from \Cref{eq:svp-gap-constants}.
Therefore \Cref{eq:svp-norm-budget} holds.
Apply the algorithm in \Cref{thm:CVP-to-SVP} to the input instance and the constructed locally dense lattice.
That theorem guarantees that both YES and NO instances are preserved and that the output parameter is $cd$.

We finally show that the total running time is $\FPT$.
The parameters $D,h,k$ depend only on $d$ and the fixed constants and are bounded by a computable function of $d$.
The time needed to construct the locally dense lattice and the size of its representation are therefore bounded by $f_{p,\gamma}(d)(m+1)^{O(1)}$ for a computable function $f_{p,\gamma}$.
The algorithm in \Cref{thm:CVP-to-SVP} takes time polynomial in the input length and the size of the constructed locally dense lattice, including the bit lengths of the integer entries.
After increasing $f_{p,\gamma}$ if necessary, the total running time is bounded by $f_{p,\gamma}(d)$ times a polynomial in the input length, with an exponent independent of $d$.
Both algorithms are deterministic, and the output parameter $cd$ depends only on $d$.
\end{proof}

Combining this reduction with the hardness of CVP proves \Cref{thm:SVP-is-W[1]hard}.

\begin{proof}[Proof of \Cref{thm:SVP-is-W[1]hard}]
We obtain the claimed hardness of SVP by composing the reductions.
Fix a computable real $p\ge1$ and a constant $1\le\gamma'<2^{1/p}$.
Choose the integer $\gamma$ from \Cref{thm:CVP-to-SVP-deterministic}.
By \Cref{thm:CVP-is-hard}, the problem $\gamma$-$\GapCVP_p$ is $\W[1]$-hard under deterministic $\FPT$ many-one reductions.
Composing these reductions with the reduction in \Cref{thm:CVP-to-SVP-deterministic} proves that $\gamma'$-$\GapSVP_p$ is $\W[1]$-hard under deterministic $\FPT$ many-one reductions.
\end{proof}
\section{A Deterministic Reduction from NCP to MDP}\label{sec:mdp}

We now show that the approach of \Cref{sec:svp} also gives the $\W[1]$-hardness of MDP under deterministic reductions.
We observe that the reduction of Hirahara and Ogitsuka~\cite[Theorem~3.5]{conf/random/HiraharaO26} becomes deterministic if we can deterministically construct a locally dense code, as defined below, in FPT time.
We work over a prime field and extend scalars at the end.

\begin{definition}[Locally dense code]\label{def:locally-dense-code}
Let $p$ be prime and let $m,k,D,r$ be positive integers.
Let $\C$ be a linear code over $\F_p$ with $N$ coordinates, let $\vec s\in\F_p^N$, and let $S$ be an ordered list of $m$ distinct coordinates.
The tuple $(\C,\vec s,S)$ is an $(m,k,D,r)$-locally dense code if
\begin{enumerate}[label=\textup{(\roman*)}]
\item\label{item:ldc-minimum-distance} $\lambda(\C)>D$;
\item\label{item:ldc-center} $\vec s|_S=\vec0$;
\item\label{item:ldc-local-density} for every $\vec u\in\F_p^m$ with $\|\vec u\|_0\le k$, there is a nonzero $\vec w\in\C-\vec s$ such that $\|\vec w\|_0\le r$ and $\vec w|_S=\vec u$.
\end{enumerate}
We represent $\C$ by a parity-check matrix over $\F_p$.
\end{definition}

\subsection{A Non-Uniform Reduction Using a Locally Dense Code}
The following reduction is essentially based on the observation that the reduction of Hirahara and Ogitsuka gives a non-uniform reduction when a locally dense code is provided as advice.
For fixed $p$, the same locally dense code can be used for all instances with the same $m$ and $k$.

\begin{theorem}[{\cite[proof of Theorem~3.5 in the full version, adapted]{conf/random/HiraharaO26}}]
\label{thm:NCP-to-MDP}
Fix a prime $p$.
There is a deterministic algorithm that, given an instance $(\mathbf G,\vec t,k)$ of $(4p)$-$\GapNCP_p$ with $\mathbf G\in\F_p^{m\times n}$ and an $(m,k,4pk,(4p-2)k)$-locally dense code $(\C,\vec s,S)$, outputs an instance of $\frac{4p}{4p-1}$-$\GapMDP_p$ with parameter $(4p-1)k$, preserving both YES and NO instances.
If the supplied parity-check matrix of $\C$ has $h$ rows and $N$ columns, the algorithm runs in time $(m+n+N+h+k)^{O(1)}$.
\end{theorem}

\begin{proof}
We first construct the output instance.
Set $D=4pk$ and $r=(4p-2)k$.
From the supplied parity-check matrix $\mathbf H$, compute a basis of $\ker\mathbf H$ by Gaussian elimination.
Use these basis vectors as the columns of $\mathbf A\in\F_p^{N\times n'}$, so that $\C(\mathbf A)=\C$.
Let $\mathbf A_S$ consist of the rows of $\mathbf A$ indexed by $S$, in the specified order.
Let $\mathbf E\in\F_p^{(D+1)m\times m}$ be the matrix that repeats each coordinate $D+1$ times.
Multiplication by $\mathbf E$ therefore multiplies the Hamming weight by $D+1$.
Output $(\mathbf G',k')$, where
\begin{equation}\label{eq:mdp-output-code}
 \mathbf G'=
 \begin{pmatrix}
  \mathbf G & \mathbf0 & -\vec t\\
  \mathbf0 & \mathbf A & -\vec s\\
  \mathbf E\mathbf G & -\mathbf E\mathbf A_S & -\mathbf E\vec t
 \end{pmatrix},
 \qquad k'=k+r=(4p-1)k.
\end{equation}

We next show completeness.
Suppose that $\|\mathbf G\vec x-\vec t\|_0\le k$ for some $\vec x\in\F_p^n$.
Since $\C(\mathbf A)=\C$, the local density condition in \Cref{def:locally-dense-code} gives $\vec y\in\F_p^{n'}$ such that
\[
 \mathbf A\vec y-\vec s\ne\vec0,\qquad
 \|\mathbf A\vec y-\vec s\|_0\le r,\qquad
 (\mathbf A\vec y-\vec s)|_S=\mathbf G\vec x-\vec t.
\]
Since $\vec s|_S=\vec0$, the last equality gives $\mathbf A_S\vec y=\mathbf G\vec x-\vec t$.
Therefore,
\[
 \mathbf G'
 \begin{pmatrix}
  \vec x\\
  \vec y\\
  1
 \end{pmatrix}
 =
 \begin{pmatrix}
  \mathbf G\vec x-\vec t\\
  \mathbf A\vec y-\vec s\\
  \vec0
 \end{pmatrix}.
\]
This codeword is nonzero because $\mathbf A\vec y-\vec s\ne\vec0$.
Its weight is at most $k+r=k'$.

We now show soundness.
Suppose that $\operatorname{dist}(\C(\mathbf G),\vec t)>D$.
Assume for contradiction that $\C(\mathbf G')$ contains a nonzero codeword of weight at most $D$.
Using $\vec s|_S=\vec0$, we can write this codeword as
\[
 \begin{pmatrix}
  \mathbf G\vec x-b\vec t\\
  \mathbf A\vec y-b\vec s\\
  \mathbf E\bigl(\mathbf G\vec x-b\vec t
                -(\mathbf A\vec y-b\vec s)|_S\bigr)
 \end{pmatrix}
\]
for some $\vec x\in\F_p^n$, $\vec y\in\F_p^{n'}$, and $b\in\F_p$.
If $b\ne0$, multiplication by $b^{-1}$ does not change Hamming weight, so the first block has weight
\[
 \|\mathbf G\vec x-b\vec t\|_0
 =\|\mathbf G(b^{-1}\vec x)-\vec t\|_0
 \ge\operatorname{dist}(\C(\mathbf G),\vec t)>D.
\]
This contradicts the assumed weight bound, so $b=0$.

The case $b=0$ also leads to a contradiction.
The second block is now $\mathbf A\vec y\in\C$ and has weight at most $D$.
Since $\lambda(\C)>D$, we must have $\mathbf A\vec y=\vec0$.
The last block is therefore $\mathbf E\mathbf G\vec x$.
Since the whole codeword is nonzero, we must have $\mathbf G\vec x\ne\vec0$.
Thus the last block has weight
\[
 \|\mathbf E\mathbf G\vec x\|_0
 =(D+1)\|\mathbf G\vec x\|_0
 \ge D+1,
\]
again contradicting the assumed weight bound.
Therefore $\lambda(\C(\mathbf G'))>D=\frac{4p}{4p-1}k'$.

We finally bound the running time.
The matrix $\mathbf G'$ has $m+N+(D+1)m$ rows and $n+n'+1$ columns, where $n'\le N$.
Since $p$ is fixed and $D=4pk$, computing $\mathbf A$ by Gaussian elimination and forming $\mathbf G'$ take $(m+n+N+h+k)^{O(1)}$ time.
\end{proof}

\subsection{A Deterministic FPT Construction of Locally Dense Codes}

In this section, we give a deterministic FPT algorithm that constructs a locally dense code as defined in \Cref{def:locally-dense-code}.
Our construction uses an extended BCH code.

\begin{theorem}\label{thm:bch-locally-dense-code}
Fix a prime $p$.
Given positive integers $k,m$, a deterministic algorithm constructs an $(m,k,4pk,(4p-2)k)$-locally dense code $(\C,\vec s,S)$.
The coordinates of $\C$ are indexed by an explicitly represented extension field $K/\F_p$, and $S=(x_1,\ldots,x_m)$ consists of distinct elements of $K$.
The algorithm outputs a parity-check matrix for $\C$, the vector $\vec s$, and the coordinates in $S$.
The field size and running time are bounded by $f_p(k)(m+1)^{O(1)}$ for a computable function $f_p$, with an exponent independent of $k$.
\end{theorem}

\begin{proof}
\setcounter{claim}{0}
We first choose the field that indexes the coordinates of the code.
Put
\[
 D=4pk,\qquad J=\{j:1\le j<D,\ p\nmid j\},\qquad
 L=|J|=4(p-1)k.
\]
Let $\Pi_J=\prod_{j\in J}j$.
Choose the smallest power $Q$ of $p$ satisfying
\[
 Q>\max\left\{D,\ m+L,\ 9\Pi_J^5,\ 9\Pi_J\bigl(1+(L+k)m\bigr)\right\},
\]
and construct $K=\F_Q$ using \Cref{lem:explicit-extension-fields}.
For $z\in K$, let $\vec e_z\in\F_p^K$ have entry $1$ at $z$ and entry $0$ at every other coordinate.

Next, we construct the output consisting of a code $\C$, a vector $\vec{s}$, and an ordered list of coordinates $S$.
We define the code $\C$ by
\begin{equation}\label{eq:bch-code}
 \C=\left\{\vec c\in\F_p^K:
    \sum_{z\in K}c_z=0,\quad
    \sum_{z\in K}c_z z^j=0\quad(j\in J)\right\}.
\end{equation}
Expanding each equation over $K$ in the chosen $\F_p$-basis of $K$ yields a parity-check matrix over $\F_p$.
To construct $\vec{s}$ and $S$, consider the matrix whose rows are indexed by $J$, columns are indexed by $y\in K$, and $(j,y)$-entry is $jy^{j-1}$.
Its rows are linearly independent over $K$.
Indeed, a nontrivial linear dependence among the rows would give a polynomial of degree at most $D-2$ that vanishes at every element of $K$.
This polynomial would be nonzero because its monomials have distinct exponents and each $j\in J$ is nonzero in $K$.
This is impossible because $D-2<Q$.
Using Gaussian elimination with the fixed enumeration of $K$, select $L$ linearly independent columns and denote their indices by $y_1,\ldots,y_L$.
Then
\begin{equation}\label{eq:bch-center-matrix}
 \left(jy_\ell^{j-1}\right)_{j\in J,\,\ell\in[L]}
 \quad\text{is invertible.}
\end{equation}
Since $Q>m+L$, we can choose the first $m$ distinct elements $x_1,\ldots,x_m$ outside $\{y_1,\ldots,y_L\}$ in the same enumeration.
Set
\[
 S=(x_1,\ldots,x_m),\qquad
 \vec s=-\sum_{\ell=1}^L\vec e_{y_\ell}.
\]
Thus $\vec s|_S=\vec0$.
When considering membership in $S$, we identify $S$ with its underlying set $\{x_1,\ldots,x_m\}$.

We prove the required minimum distance of $\C$.
\begin{claim}\label{clm:bch-distance}
The code in \Cref{eq:bch-code} satisfies $\lambda(\C)>D$.
\end{claim}
\begin{proof}[Proof of \Cref{clm:bch-distance}]
Suppose for contradiction that there exists a nonzero codeword $\vec c\in\C \setminus \{\vec{0}\}$ with support $\{z_1,\ldots,z_t\}$, where $1\le t\le D$.
For each $1\le b<t$, write $b=p^e j$ with $p\nmid j$.
Since $j\le b<D$, we have $j\in J$.
By \Cref{eq:bch-code}, we have
\[
 \sum_{i=1}^t c_{z_i}z_i^j=0.
\]
Taking the $p^e$-th power of both sides gives
\[
 \sum_{i=1}^t c_{z_i}z_i^{jp^e}
 =\sum_{i=1}^t c_{z_i}z_i^b=0,
\]
since $K$ has characteristic $p$ and $c_{z_i}\in\F_p$.
Together with $\sum_{i=1}^t c_{z_i}=0$, these equations can be written as
\[
 \begin{pmatrix}
  1 & 1 & \cdots & 1\\
  z_1 & z_2 & \cdots & z_t\\
  \vdots & \vdots & \ddots & \vdots\\
  z_1^{t-1} & z_2^{t-1} & \cdots & z_t^{t-1}
 \end{pmatrix}
 \begin{pmatrix}
  c_{z_1}\\
  c_{z_2}\\
  \vdots\\
  c_{z_t}
 \end{pmatrix}
 =\vec0.
\]
The Vandermonde matrix is invertible because the $z_i$ are distinct.
Thus every $c_{z_i}$ is zero, contrary to the choice of $\vec c$.
\end{proof}

We now construct, for each $\vec u\in\F_p^m$ with $\|\vec u\|_0\le k$, a vector $\vec w\in\C-\vec s$ satisfying
\[
 \vec w|_S=\vec u,\qquad
 0<\|\vec w\|_0\le(4p-2)k,
\]
as required by condition~\ref{item:ldc-local-density} of \Cref{def:locally-dense-code}.
If $\vec u=\vec0$, choose $\vec w=-\vec s$.
Then $\vec w+\vec s=\vec0\in\C$, $\vec w|_S=\vec0$, and $0<\|\vec w\|_0=L\le(4p-2)k$.
For the rest of the proof, fix $\vec u\ne\vec0$ and put $T=\{i\in[m]:u_i\ne0\}$, so that $1\le|T|\le k$.
For any $z_1,\ldots,z_L,\tau\in K$, define
\begin{equation}\label{eq:bch-completion-vector}
 \vec w=\sum_{\ell=1}^L\vec e_{z_\ell}
       +\sum_{i\in T}u_i(\vec e_{x_i}-\vec e_{x_i+\tau}).
\end{equation}
The support of $\vec w$ is contained in the set of indices appearing in this expression.
Hence, for every choice of these field elements,
\[
 \|\vec w\|_0\le L+2|T|\le L+2k=(4p-2)k.
\]

We derive the equations equivalent to $\vec w\in\C-\vec s$.
By \Cref{eq:bch-code}, the first condition for $\vec w+\vec s\in\C$ is
\[
 \sum_{z\in K}(w_z+s_z)=0.
\]
This condition holds for every choice of $z_1,\ldots,z_L,\tau\in K$, since
\[
 \sum_{z\in K}(w_z+s_z)
 =\sum_{\ell=1}^L(1-1)+\sum_{i\in T}u_i(1-1)=0.
\]
By \Cref{eq:bch-code}, the remaining conditions for $\vec w+\vec s\in\C$ are
\[
 \sum_{z\in K}(w_z+s_z)z^j=0
 \qquad(j\in J).
\]
For each $j\in J$ and $\ell\in[L]$, we have
\[
 \sum_{z\in K}(\vec e_{z_\ell})_z z^j=z_\ell^j,
\]
since $\vec e_{z_\ell}$ has entry $1$ at $z_\ell$ and entry $0$ at every other coordinate.
Applying the same calculation to each term of $\vec w+\vec s$ gives
\[
 \begin{aligned}
 \sum_{z\in K}(w_z+s_z)z^j
 &=\sum_{\ell=1}^L(z_\ell^j-y_\ell^j)
   +\sum_{i\in T}u_i\bigl(x_i^j-(x_i+\tau)^j\bigr)\\
 &=\sum_{\ell=1}^L(z_\ell^j-y_\ell^j)
   -\sum_{i\in T}u_i\bigl((x_i+\tau)^j-x_i^j\bigr).
 \end{aligned}
\]
Let $Z_1,\ldots,Z_L,Y$ be variables over $K$.
Therefore, $\vec w+\vec s\in\C$ if and only if $(z_1,\ldots,z_L,\tau)$ satisfies
\begin{equation}\label{eq:bch-completion-equations}
 \sum_{\ell=1}^L(Z_\ell^j-y_\ell^j)
 -\sum_{i\in T}u_i\bigl((x_i+Y)^j-x_i^j\bigr)=0
 \qquad(j\in J).
\end{equation}

To ensure $\vec w|_S=\vec u$, we introduce a polynomial $g$ such that $g(z_1,\ldots,z_L,\tau)\ne0$ implies $z_\ell\notin S$ for every $\ell\in[L]$ and $x_i+\tau\notin S$ for every $i\in T$.
Set
\[
  E=\{x_b-x_i:i\in T,\ b\in[m]\}.
\]
Since $T\ne\varnothing$, we have $0\in E$ and $|E|\le km$.
Define the polynomial $g\in K[Z_1,\ldots,Z_L,Y]$ by
\begin{equation}\label{eq:bch-g}
 g(Z_1,\ldots,Z_L,Y)
 =Y\prod_{e\in E\setminus\{0\}}(Y-e)
   \prod_{\ell=1}^L\prod_{b=1}^m(Z_\ell-x_b).
\end{equation}
An empty product is $1$.
\begin{claim}\label{clm:bch-restriction}
For every $z_1,\ldots,z_L,\tau\in K$ with $g(z_1,\ldots,z_L,\tau)\ne0$, we have
\[
  z_\ell\notin S\quad(\ell\in[L]),\qquad
  x_i+\tau\notin S\quad(i\in T).
\]
The vector in \Cref{eq:bch-completion-vector} consequently satisfies $\vec w|_S=\vec u$ and $\vec w\ne\vec0$.
\end{claim}
\begin{proof}[Proof of \Cref{clm:bch-restriction}]
We first show that $z_\ell\notin S$ for every $\ell\in[L]$.
Since $g(z_1,\ldots,z_L,\tau)\ne0$, every factor of $g$ is nonzero at $(z_1,\ldots,z_L,\tau)$.
The factors $Z_\ell-x_b$ therefore give $z_\ell\ne x_b$ for every $\ell\in[L]$ and $b\in[m]$.
Since $S$ consists of $x_1,\ldots,x_m$, we have $z_\ell\notin S$ for every $\ell\in[L]$.

We next show that $x_i+\tau\notin S$ for every $i\in T$.
The factor $Y$ gives $\tau\ne0$, and the factors $Y-e$ give $\tau\ne e$ for every $e\in E\setminus\{0\}$.
Thus $\tau\notin E$.
For every $i\in T$ and $b\in[m]$, the definition of $E$ gives $x_b-x_i\in E$.
Hence $\tau\ne x_b-x_i$, which implies $x_i+\tau\ne x_b$.
Therefore $x_i+\tau\notin S$ for every $i\in T$.

We now show that $\vec w|_S=\vec u$ and $\vec w\ne\vec0$.
The vectors $\vec e_{z_\ell}$ and $\vec e_{x_i+\tau}$ have entry $0$ at every coordinate in $S$, as their indices lie outside $S$.
By \Cref{eq:bch-completion-vector}, the entries of $\vec w$ on $S$ are therefore the same as those of $\sum_{i\in T}u_i\vec e_{x_i}$.
Since $x_1,\ldots,x_m$ are distinct, this gives $w_{x_b}=u_b$ for every $b\in T$.
For $b\in[m]\setminus T$, the sum has entry $0$ at $x_b$, and $u_b=0$ by the definition of $T$.
Thus $w_{x_b}=u_b$ for every $b\in[m]$, which proves $\vec w|_S=\vec u$.
Since $\vec u\ne\vec0$, we also have $\vec w\ne\vec0$.
\end{proof}

To complete the construction of $\vec w$, it is enough to show that there exists a point $(z_1,\ldots,z_L,\tau)\in K^{L+1}$ satisfying \Cref{eq:bch-completion-equations} and $g(z_1,\ldots,z_L,\tau)\ne0$, where $g$ is defined in \Cref{eq:bch-g}.
For such a point, the vector $\vec w$ defined in \Cref{eq:bch-completion-vector} belongs to $\C-\vec s$ and satisfies $\|\vec w\|_0\le(4p-2)k$.
By \Cref{clm:bch-restriction}, it also satisfies $\vec w|_S=\vec u$ and $\vec w\ne\vec0$.
We prove that such a point exists by applying \Cref{lem:keep-all-but-one}.

We first show that $\xi=(y_1,\ldots,y_L,0)$ is a common zero of the left sides of \Cref{eq:bch-completion-equations} and $g$.
At $\xi$, we have $Z_\ell=y_\ell$ for every $\ell\in[L]$ and $Y=0$.
Thus every left side in \Cref{eq:bch-completion-equations} is zero, since both $Z_\ell^j-y_\ell^j$ and $(x_i+Y)^j-x_i^j$ are zero at this point.
Since $g$ has the factor $Y$, we also have $g(\xi)=0$.

We next show that the Jacobian matrix of these $L+1$ polynomials is invertible at $\xi$.
We order the $L$ left sides by increasing $j\in J$, followed by $g$, and use the variable order $(Z_1,\ldots,Z_L,Y)$.
For the left side indexed by $j$, the derivative with respect to $Z_\ell$ is $jZ_\ell^{j-1}$, whose value at $\xi$ is $jy_\ell^{j-1}$.
Thus the upper left block is the invertible matrix in \Cref{eq:bch-center-matrix}.
Each derivative of $g$ with respect to $Z_\ell$ still has the factor $Y$, so
\[
 \frac{\partial g}{\partial Z_\ell}(\xi)=0
 \qquad(\ell\in[L]).
\]
When differentiating $g$ with respect to $Y$, every term except the one obtained by differentiating the first factor $Y$ still contains $Y$ and is zero at $\xi$.
Therefore,
\[
 \frac{\partial g}{\partial Y}(\xi)
 =\prod_{e\in E\setminus\{0\}}(-e)
  \prod_{\ell=1}^L\prod_{b=1}^m(y_\ell-x_b)
 \ne0,
\]
since $e\ne0$ and $y_\ell\ne x_b$.
The Jacobian matrix thus has an invertible upper left block, a zero lower left block, and a nonzero lower right entry, so it is invertible.

We now check the remaining assumptions of \Cref{lem:keep-all-but-one} concerning the degrees and the size of $K$.
For each $j\in J$, every term on the left side of \Cref{eq:bch-completion-equations} has degree at most $j$, and the coefficient of $Z_1^j$ is $1$.
Its degree is therefore exactly $j$, and the product of these degrees is $\Pi_J$.
Since $g$ is a product of linear factors, we have
\[
 \deg g=1+|E\setminus\{0\}|+Lm\le1+(L+k)m.
\]
All $L+1$ polynomials are nonconstant.
Our choice of $Q$ gives
\[
 Q>9\Pi_J^5,\qquad
 Q>9\Pi_J\bigl(1+(L+k)m\bigr)\ge9\Pi_J\deg g.
\]

We can now apply \Cref{lem:keep-all-but-one} to obtain the required point.
We use the $L$ left sides of \Cref{eq:bch-completion-equations}, followed by $g$, as the polynomials in the lemma, with variables $(Z_1,\ldots,Z_L,Y)$.
In this application, $n=L+1\ge2$, $\Delta=\Pi_J$, and $f_n=g$.
The lemma gives a point $(z_1,\ldots,z_L,\tau)\in K^{L+1}$ satisfying \Cref{eq:bch-completion-equations} and $g(z_1,\ldots,z_L,\tau)\ne0$.

The point given by the lemma completes the construction of $\vec w$ for the fixed nonzero vector $\vec u$.
Together with the case $\vec u=\vec0$ treated above, this proves condition~\ref{item:ldc-local-density} of \Cref{def:locally-dense-code}.
\Cref{clm:bch-distance} and $\vec s|_S=\vec0$ give the other two conditions.

We finally bound the running time and the size of the output.
The algorithm constructs $K$, computes the parity-check matrix and the matrix used to select $y_1,\ldots,y_L$, and outputs the parity-check matrix, $S$, and $\vec s$.
Since $Q$ is the smallest power of $p$ satisfying the required inequality, we have
\[
 Q\le p\max\left\{D,\ m+L,\ 9\Pi_J^5,\
                  9\Pi_J\bigl(1+(L+k)m\bigr)\right\}.
\]
Since $D$, $L$, and $\Pi_J$ depend only on $p$ and $k$, this gives $Q\le f_p(k)(m+1)$ for a computable function $f_p$.
The parity-check matrix has at most $1+L\log_p Q$ rows and $Q$ columns.
The matrix used to select $y_1,\ldots,y_L$ has $L$ rows and $Q$ columns.
Computing these matrices and selecting the coordinates require a number of field operations polynomial in $Q$, since $D,L,m<Q$.
By \Cref{lem:explicit-extension-fields}, constructing $K$, enumerating its elements, and performing these operations take time polynomial in $Q$.
The output size is also polynomial in $Q$.
Therefore, after increasing $f_p$ if necessary, both the running time and the output size are at most $f_p(k)(m+1)^{O(1)}$, where the exponent does not depend on $k$.
\end{proof}

\subsection{Finishing the Proof}

We first obtain a deterministic $\FPT$ reduction from NCP to MDP over prime fields.
The reduction in \Cref{thm:NCP-to-MDP} takes a locally dense code as additional input, and \Cref{thm:bch-locally-dense-code} constructs such a code deterministically in $\FPT$ time.

\begin{theorem}\label{thm:NCP-to-MDP-someapprox}
For every fixed prime $p$, there is a deterministic $\FPT$ many-one reduction from $(4p)$-$\GapNCP_p$ to $\frac{4p}{4p-1}$-$\GapMDP_p$.
The output parameter is $(4p-1)k$, where $k$ is the input parameter.
\end{theorem}
\begin{proof}
We construct the reduction by applying the two algorithms in sequence.
Given an instance $(\mathbf G,\vec t,k)$ of $(4p)$-$\GapNCP_p$ with $\mathbf G\in\F_p^{m\times n}$, use \Cref{thm:bch-locally-dense-code} to construct an $(m,k,4pk,(4p-2)k)$-locally dense code $(\C,\vec s,S)$.
Apply the algorithm in \Cref{thm:NCP-to-MDP} to $(\mathbf G,\vec t,k)$ and $(\C,\vec s,S)$.
That theorem guarantees that both YES and NO instances are preserved and that the output parameter is $(4p-1)k$.

We next show that the total running time is $\FPT$.
The time needed to construct the locally dense code and the size of its representation are bounded by $f_p(k)(m+1)^{O(1)}$ for a computable function $f_p$.
The algorithm in \Cref{thm:NCP-to-MDP} takes time polynomial in $m+n+k$ and the size of the constructed locally dense code.
Thus, after increasing $f_p$ if necessary, the total running time is bounded by $f_p(k)(m+n)^{O(1)}$, with an exponent independent of $k$.
Both algorithms are deterministic, and for fixed $p$ the output parameter $(4p-1)k$ depends only on $k$.
\end{proof}

To transfer hardness from $\F_p$ to $\F_q$ with $q=p^a$, we view the same generator matrix over $\F_q$.
The following lemma shows that this preserves the minimum distance and hence both YES and NO instances of MDP.

\begin{lemma}\label{lem:scalar-extension-distance}
Let $q=p^a$, where $p$ is prime and $a\ge1$ is an integer, and let $\mathbf G$ be a matrix over $\F_p$.
The code generated by the columns of $\mathbf G$ has the same minimum distance whether the linear combinations are taken over $\F_p$ or over $\F_q$.
\end{lemma}
\begin{proof}
Allowing coefficients from $\F_q$ cannot increase the minimum distance.
Indeed, every linear combination of the columns of $\mathbf G$ with coefficients in $\F_p$ is also one with coefficients in $\F_q$.

To prove the reverse inequality, we show that every nonzero codeword over $\F_q$ gives a nonzero codeword over $\F_p$ of no greater weight.
Express a nonzero codeword over $\F_q$ as a linear combination of the columns of $\mathbf G$.
Expand the coefficients of this linear combination in an $\F_p$-basis of $\F_q$.
Since $\mathbf G$ has entries in $\F_p$, collecting the terms for each basis element gives a codeword over $\F_p$.
At least one of these codewords is nonzero, since the original codeword is nonzero.
If a coordinate of the original codeword is zero, linear independence of the basis implies that the same coordinate is zero in each of these codewords.
Thus each has support contained in that of the original codeword, and a nonzero one has no greater weight.
This proves the reverse inequality.
If $\mathbf G$ generates the zero code, it does so over both fields, and both minimum distances are $+\infty$ by our convention.
\end{proof}

To complete the proof of \Cref{thm:MDP-is-W[1]hard}, we use tensor powers to obtain hardness for every constant approximation factor.
This step uses the fact that the minimum distance of a tensor product of two nonzero linear codes is the product of their minimum distances.

\begin{proof}[Proof of \Cref{thm:MDP-is-W[1]hard}]
We first establish hardness for the factor $4p/(4p-1)$ over the fixed field $\F_q$, where $q=p^a$ and $p$ is prime.
By \Cref{thm:NCP-is-hard,thm:NCP-to-MDP-someapprox}, the problem $\frac{4p}{4p-1}$-$\GapMDP_p$ is $\W[1]$-hard under deterministic $\FPT$ many-one reductions.
By \Cref{lem:scalar-extension-distance}, viewing the same generator matrix over $\F_q$ preserves the minimum distance.
This step is deterministic, takes polynomial time, and leaves the parameter unchanged.
Thus the same hardness result holds over $\F_q$.

We next prove the formula for the minimum distance of a tensor product.
View a nonzero codeword in the tensor product of two nonzero linear codes as a matrix whose columns belong to the first code and whose rows belong to the second.
At least one column is nonzero, so the number of nonzero rows is at least the minimum distance of the first code.
Each nonzero row has weight at least the minimum distance of the second code.
The total weight is therefore at least the product of the two minimum distances.
For the reverse inequality, choose a codeword of minimum weight from each code and take their tensor product.
Its weight is exactly the product of the two minimum distances.

We now construct a reduction to $\gamma$-$\GapMDP_q$ for any fixed $\gamma\ge1$.
Since $4p/(4p-1)>1$, we can choose a positive integer $\ell$ such that
\[
 \left(\frac{4p}{4p-1}\right)^\ell\ge\gamma.
\]
Map an instance $(\mathbf G,k)$ of $\frac{4p}{4p-1}$-$\GapMDP_q$ to $(\mathbf G^{\otimes\ell},k^\ell)$.

We show that this map preserves both YES and NO instances.
Repeatedly applying the formula for the minimum distance of a tensor product shows that, for a YES instance,
\[
 \lambda\bigl(\C(\mathbf G^{\otimes\ell})\bigr)
 =\lambda\bigl(\C(\mathbf G)\bigr)^\ell
 \le k^\ell.
\]
For a NO instance whose code is nonzero, we have
\[
 \lambda\bigl(\C(\mathbf G^{\otimes\ell})\bigr)
 =\lambda\bigl(\C(\mathbf G)\bigr)^\ell
 >
 \left(\frac{4p}{4p-1}\right)^\ell k^\ell
 \ge\gamma k^\ell.
\]
A zero input code remains zero and is therefore also mapped to a NO instance.

We finally check the running time and the output parameter.
If $\mathbf G$ has $m$ rows and $n$ columns, then $\mathbf G^{\otimes\ell}$ has $m^\ell$ rows and $n^\ell$ columns.
Since $\ell$ depends only on the fixed constants $p$ and $\gamma$, the output matrix and parameter can be computed deterministically in polynomial time.
The new parameter is $k^\ell$, which depends only on $k$.
Combining this reduction with the hardness result established above proves the theorem.
\end{proof}

\section*{AI Methodology}
The deterministic constructions of locally dense codes and locally dense lattices were discovered by GPT-5.5 in response to prompts based on ideas for obtaining such constructions, building on the results in Sections 3.1 and 4.1.
Initially, the constructions and their proofs relied on complex concepts from algebraic geometry.
The authors subsequently obtained simpler constructions and proofs with the assistance of GPT-5.6 and GPT-6.
The authors verified the correctness and originality of all content, including references.

\printbibliography
\appendix
\section{\texorpdfstring{$\W[1]$}{W[1]}-hardness of SVP for any constant approximation factor}\label{sec:svp-extend}
\setcounter{claim}{0}

In this section, we show that SVP is $\W[1]$-hard to approximate within every constant factor for every $p\ge1$.
We reduce from the Maximum Likelihood Decoding Problem over a prime field, in two steps.
First, we embed an MLD instance into a lattice using the block construction of \Cref{thm:CVP-to-SVP}.
This produces a gap of roughly $2$ between YES and NO instances, together with a lower bound on the minimum distance of the lattice modulo the prime.
Second, we amplify the gap by taking tensor powers, following Wan~\cite{wan2026anyconstant}.

MLD is the parity-check formulation of NCP.
For \(p>1\), Bennett, Cheraghchi, Guruswami, and Ribeiro~\cite{journals/sicomp/BennettCGR24} use a randomized reduction from NCP over \(\mathbb{F}_2\) to SVP in the \(\ell_p\) norm and then take tensor powers to obtain hardness for every constant approximation factor.
In the following definition, we additionally require that a YES instance has a binary solution.
\begin{definition}[$\gamma$-$\GapMLD_q$, \cite{journals/BhattacharyyaBE+21}]
\label{def:svp-ext-mld}
For $\gamma\geq 1$ and a prime $q$, the parameterized promise problem $\gamma$-approximate Maximum Likelihood Decoding Problem (MLD) over $\F_q$ ($\gamma$-$\GapMLD_q$) is defined as follows.
Instances are tuples $(\mathbf{M},\vec{t},k)$, where $\mathbf{M}\in\F_q^{r\times m}$ is a parity-check matrix of $\ker\mathbf{M}$, $\vec{t}\in\F_q^r$ is a target vector, and $k\in\mathbb{Z}_{>0}$ is a distance parameter such that
\begin{enumerate}
    \item $(\mathbf{M},\vec{t},k)$ is a YES instance if there exists $\vec{x}\in\{0,1\}^m$ such that $\mathbf{M}\vec{x}=\vec{t}$ and $\|\vec{x}\|_0\leq k$;
    \item $(\mathbf{M},\vec{t},k)$ is a NO instance if $\|\vec{x}\|_0>\gamma k$ for all $\vec{x}\in\F_q^m$ with $\mathbf{M}\vec{x}=\vec{t}$.
\end{enumerate}
The parameter of interest is $k$.
\end{definition}

\begin{theorem}[{\cite[Theorems~4.2, 4.9, and~5.4, adapted]{journals/BhattacharyyaBE+21}}]
\label{thm:svp-ext-mld-hard}
Fix a constant $\gamma\ge1$, and let $k\mapsto q(k)$ be any computable function whose values are prime.
Then $\gamma$-$\GapMLD_{q(k)}$ is $\W[1]$-hard under deterministic $\FPT$ many-one reductions.
\end{theorem}

The reductions in~\cite{journals/BhattacharyyaBE+21} produce YES instances with binary solutions.
Their output parameter is determined before the field is chosen, so the prime may be any computable function of this parameter without affecting the $\FPT$ running time.

In our reduction, the prime $q$ is fixed in advance as a function of the parameter $k$.
The construction in \Cref{thm:rs-locally-dense-lattice}, however, chooses a prime that grows with $m$.
We therefore index the coordinates of the Reed--Solomon lattice by an extension field of $\F_q$.

\begin{theorem}\label{thm:appendix-rs-locally-dense-lattice}
Let $m,h\ge 1$ and $k\ge 2$ be integers, and let $q>2k$ be a prime.
There is a deterministic algorithm that, given $m,h,k,q$, runs in time $f(q+h+k)\cdot(m+1)^{O(1)}$ for a computable function $f$ and outputs a matrix $\mathbf{H}\in\F_q^{r\times N}$, a vector $\vec{s}\in\Z^N$, and an ordered list $S$ of $m$ distinct coordinates.
The output satisfies the following conditions, where $L=\{\vec{v}\in\Z^N:\mathbf{H}(\vec{v}\bmod q)=\vec{0}\}$.
\begin{enumerate}[label=\textup{(\roman*)}]
\item\label{item:app-distance} $\lambda_1^{(1)}(L)\ge 2k$ and $\lambda(L\bmod q)\ge k+1$.
\item\label{item:app-center} $\vec{s}|_S=\vec{0}$.
\item\label{item:app-density} For every $\vec{u}\in\Z^m$ with $\|\vec{u}\|_0\le h$, there exists $\vec{w}\in L-\vec{s}$ with $\vec{w}|_S=\vec{u}$ such that
\[
  \|\vec{w}\|_p^p=k-1+2\|\vec{u}\|_p^p
  \qquad\text{for every real }p\ge 1.
\]
\end{enumerate}
The output of \Cref{thm:rs-locally-dense-lattice} also satisfies \ref{item:app-distance}--\ref{item:app-density}.
\end{theorem}

\begin{proof}
Let $T$ be the integer defined in \Cref{eq:svp-prime-bound} with $L$ replaced by $k-1$ and $D$ replaced by $2$.
Since the maximum in \Cref{eq:svp-prime-bound} contains $2k>2$, the integer $T$ depends only on $m$, $h$, and $k$.
For the same reason, $T$ equals the integer used in the proof of \Cref{thm:rs-locally-dense-lattice} whenever $D^p<2k$, since then $D\le D^p<2k$.

We first describe the construction.
Let $a\ge1$ be the least integer such that $Q := q^a>T$, and set $N=Q$.
Set $L:=L_{Q,k}$ as defined in \Cref{eq:svp-rs-lattice}, with coordinates indexed by $\F_Q$.
We construct a parity-check matrix $\mathbf{H}$ for $L\bmod q$ over $\F_q$ as follows.
Fix an $\F_q$-basis of $\F_Q$ and replace each row $(z^j)_{z\in\F_Q}$, for $0\le j<k$, by the $a$ rows of its coordinates in this basis.
Here $z^0=1$ for every $z\in\F_Q$.
The resulting matrix $\mathbf{H}$ has $ka$ rows and $N$ columns.
By \Cref{eq:svp-rs-lattice}, an integer vector $\vec{v}$ belongs to $L$ if and only if $\mathbf{H}(\vec{v}\bmod q)=\vec{0}$.
Since $Q>T>m+k-1$, we can choose distinct elements $y_1,\ldots,y_{k-1},x_1,\ldots,x_m\in\F_Q$ in a fixed enumeration of $\F_Q$.
Set
\[
 \vec{s}=-\sum_{\ell=1}^{k-1}\vec{e}_{y_\ell},\qquad S=(x_1,\ldots,x_m).
\]
The algorithm outputs $\mathbf{H}$, $\vec{s}$, and $S$.

Conditions~\ref{item:app-distance} and~\ref{item:app-center} follow directly from the construction.
Condition~\ref{item:app-distance} is \Cref{lem:svp-rs-lattice} applied to $Q=q^a$, whose hypotheses hold because $k\ge2$ and $q>2k$.
Condition~\ref{item:app-center} holds because the elements $y_\ell$ and $x_i$ are distinct.

We now prove condition~\ref{item:app-density}.
Fix $\vec{u}\in\Z^m$ with $\|\vec{u}\|_0\le h$, and put $I=\{i\in[m]:u_i\ne0\}$.
We construct $\vec{w}\in L-\vec{s}$ with $\vec{w}|_S=\vec{u}$ whose nonzero entries are exactly $k-1$ entries equal to $1$ together with $u_i$ and $-u_i$ for each $i\in I$.
Such a vector satisfies $\|\vec{w}\|_p^p=k-1+2\sum_{i\in I}|u_i|^p=k-1+2\|\vec{u}\|_p^p$ for every real $p\ge1$.
If $I=\varnothing$, then $\vec{w}=-\vec{s}$ has the required form.

Suppose that $I\ne\varnothing$.
We repeat the argument in the proof of \Cref{thm:rs-locally-dense-lattice} with $\F_Q$ in place of $\F_q$ and $k-1$ in place of $L$.
For $z_1,\ldots,z_{k-1},\tau\in\F_Q$, define $\vec{w}$ by \Cref{eq:svp-completion-vector}.
The computation in that proof shows that $\vec{w}\in L-\vec{s}$ if and only if $(z_1,\ldots,z_{k-1},\tau)$ satisfies \Cref{eq:svp-completion-equations} over $\F_Q$.
Define $g$ by \Cref{eq:svp-last-polynomial} over $\F_Q$.
By the argument of \Cref{clm:rs-restriction}, if $g(z_1,\ldots,z_{k-1},\tau)\ne0$, then all coordinate indices in \Cref{eq:svp-completion-vector} are pairwise distinct and the indices $z_\ell$ and $x_i+\tau$ lie outside $S$.
In this case, $\vec{w}|_S=\vec{u}$ and the nonzero entries of $\vec{w}$ are as described above.

It remains to find a point satisfying \Cref{eq:svp-completion-equations} at which $g$ is nonzero.
We apply \Cref{lem:keep-all-but-one} over $\F_Q$ to the $k-1$ left sides of \Cref{eq:svp-completion-equations} followed by $g$, exactly as in the proof of \Cref{thm:rs-locally-dense-lattice}.
The point $(y_1,\ldots,y_{k-1},0)$ is a common zero of these $k\ge2$ polynomials.
The Jacobian matrix is invertible at this point, because $\F_Q$ has characteristic $q>k-1$ and the elements $y_\ell$ and $x_i$ are distinct.
The product of the degrees of the first $k-1$ polynomials is $\Delta=(k-1)!$.
The bound on $\deg g$ in that proof uses only $|I|\le h$, so it holds here as well.
Since $Q>T$, we therefore have $Q>9\Delta^5$ and $Q>9\Delta\deg g$.
\Cref{lem:keep-all-but-one} gives the required point.

We next show that the output of \Cref{thm:rs-locally-dense-lattice} also satisfies conditions~\ref{item:app-distance}--\ref{item:app-density}.
The construction there chooses a prime $q$ with $T<q<2T$.
For this $q$, the least $a$ with $q^a>T$ is $a=1$.
The construction then defines $\mathbf{H}$, $\vec{s}$, and $S$ exactly as above.
Thus its output coincides with the output of our algorithm on input $(m,h,k,q)$.

We finally bound the running time.
By \Cref{eq:svp-prime-bound}, we have $T\le f_0(h+k)(m+1)$ for a computable function $f_0$.
The minimality of $a$ gives $q^{a-1}\le T$, so $Q\le qT$.
We construct $\F_Q=\F_q[X]/(g_0)$ for a monic irreducible $g_0\in\F_q[X]$ of degree $a$ by the search in the proof of \Cref{lem:explicit-extension-fields}.
Here $q$ is part of the input, but each division in that search takes time polynomial in $a$ and $\log q$, so the search takes time polynomial in $Q$.
Using the $\F_q$-basis $1,X,\ldots,X^{a-1}$, we compute all entries of $\mathbf{H}$ in time polynomial in $Q$ and $k$.
Choosing $\vec{s}$ and $S$ also takes time polynomial in $Q$.
The point given by \Cref{lem:keep-all-but-one} is used only in the analysis and is not computed.
Thus the running time and the output size are polynomial in $Q+k\le qf_0(h+k)(m+1)+k$, and hence bounded by $f(q+h+k)(m+1)^{O(1)}$ for a computable function $f$.
\end{proof}

In particular, for every real $p\ge 1$ and all integers $d\ge 1$ and $D\ge 2$ with $d^p\le h$ and $D^p<2k$, the tuple $(q,\mathbf{H},\vec{s},S)$ is a locally dense lattice with parameters $(p,m,d,D,k-1+2h)$.
Indeed, $\|\vec{u}\|_p\le d$ implies $\|\vec{u}\|_0\le\|\vec{u}\|_p^p\le h$, and $\|\vec{v}\|_p^p\ge\|\vec{v}\|_1\ge 2k>D^p$ for every nonzero $\vec{v}\in L$.

We now turn to tensor products.
For integer lattices $U=\lat(\mathbf{B}_1)$ and $V=\lat(\mathbf{B}_2)$, the tensor product $U\otimes V$ is the lattice generated by $\mathbf{B}_1\otimes\mathbf{B}_2$.
Equivalently, it is the lattice generated by the vectors $\vec{u}\otimes\vec{v}$ with $\vec{u}\in U$ and $\vec{v}\in V$, so it does not depend on the choice of bases.
We write $U^{\otimes j}$ for the tensor product of $j$ copies of $U$.
If the columns of $\mathbf{B}_1$ and $\mathbf{B}_2$ are linearly independent, then so are those of $\mathbf{B}_1\otimes\mathbf{B}_2$.
We use the following two lemmas from the second version of~\cite{wan2026anyconstant}.

\begin{lemma}[{\cite[Lemma~2.1]{wan2026anyconstant}}]\label{lem:svp-ext-code-tensor}
Let $\mathcal{C}_1$ and $\mathcal{C}_2$ be linear codes over $\F_q$.
If $\lambda(\mathcal{C}_1)\ge D_1$ and $\lambda(\mathcal{C}_2)\ge D_2$, then $\lambda(\mathcal{C}_1\otimes\mathcal{C}_2)\ge D_1D_2$.
\end{lemma}

\begin{lemma}[{\cite[Lemma~4.2]{wan2026anyconstant}}]\label{lem:svp-ext-rank-one}
Let $U$ and $V$ be integer lattices, and write a nonzero vector of $U\otimes V$ as a matrix $\mathbf{X}$ with rows indexed by the coordinates of $U$ and columns indexed by the coordinates of $V$.
If $\mathbf{X}$ has rank $r$ over the reals, there are linearly independent vectors $\vec{u}_1,\ldots,\vec{u}_r\in U$ and linearly independent vectors $\vec{v}_1,\ldots,\vec{v}_r\in V$ such that
\[
 \mathbf{X}=\sum_{i=1}^r\vec{u}_i\vec{v}_i^{\mathsf{T}}.
\]
\end{lemma}

The following amplification lemma improves the argument of \cite[Section~5]{wan2026anyconstant} to the $\ell_1$ norm.
Since $\|\vec{z}\|_p^p\ge\|\vec{z}\|_1$ for every integer vector $\vec{z}$, the $\ell_1$ bound yields a bound for every $p\ge1$.

\begin{lemma}\label{lem:svp-ext-amplification}
For every integer $b\ge4$, one can compute an even integer $T\ge2$ such that the following holds.
Let $L$ be a nonzero integer lattice, let $K\ge2$ be an integer, and let $q>2(bK^T)^2$ be prime.
If $\lambda_1^{(1)}(L)\ge2K$ and $\lambda(L\bmod q)\ge K$, then for every real $p\ge1$,
\[
 \bigl(\lambda_1^{(p)}(L^{\otimes T})\bigr)^p\ge bK^T.
\]
\end{lemma}

\begin{proof}
Let $\eta=1/(2(8b+1)^2)$, and let $T$ be the least even integer $T\ge2$ such that $2(1+\eta)^{T-1}\ge b$.
For $1\le j\le T$, let $a_j=\min\{b,2(1+\eta)^{j-1}\}$.
Then $a_1=2$, $a_T=b$, and $a_{j+1}\le(1+\eta)a_j\le2a_j$ for $1\le j<T$.
Since $\|\vec{z}\|_p^p\ge\|\vec{z}\|_1$ for every integer vector $\vec{z}$, it suffices to prove that
\begin{equation}\label{eq:svp-ext-tensor-induction}
 \lambda_1^{(1)}(L^{\otimes j})\ge a_jK^j
 \qquad(1\le j\le T),
\end{equation}
because the case $j=T$ is the claim.
We proceed by induction on $j$.
The case $j=1$ is the assumption $\lambda_1^{(1)}(L)\ge2K$.

Before the induction step, we relate the codes obtained from $L$ and its tensor powers.
Let $\mathbf{B}$ be a basis of $L$, and let $\mathcal{C}=L\bmod q$, which is the code generated by $\mathbf{B}\bmod q$.
Since $L^{\otimes j}$ is generated by $\mathbf{B}^{\otimes j}$ and $\mathbf{B}^{\otimes j}\bmod q=(\mathbf{B}\bmod q)^{\otimes j}$, we have
\[
 L^{\otimes j}\bmod q=\mathcal{C}^{\otimes j}.
\]
Since $\lambda(\mathcal{C})\ge K$, repeated application of \Cref{lem:svp-ext-code-tensor} gives $\lambda(\mathcal{C}^{\otimes j})\ge K^j$.

Fix $1\le j<T$ and assume \Cref{eq:svp-ext-tensor-induction} for $j$.
Suppose for contradiction that some nonzero $\vec{x}\in L^{\otimes(j+1)}=L\otimes L^{\otimes j}$ satisfies $\|\vec{x}\|_1<a_{j+1}K^{j+1}$.
Write $L\subseteq\Z^N$, and view $\vec{x}$ as an $N\times N^j$ matrix $\mathbf{X}$ whose rows are indexed by the coordinates of the first factor $L$ and whose columns are indexed by the coordinates of $L^{\otimes j}$.
By the definition of tensor product, $\vec{x}$ is an integer combination of tensors of the form $\vec{u}\vec{v}^{\mathsf{T}}$, where $\vec{u}\in L$ and $\vec{v}\in L^{\otimes j}$, therefore every column of $\mathbf{X}$ lies in $L$ and the transpose of every row lies in $L^{\otimes j}$.
Writing $\|\mathbf{X}\|_1$ for the sum of the absolute values of the entries, we have
\begin{equation}\label{eq:svp-ext-short-tensor}
 \|\mathbf{X}\|_1=\|\vec{x}\|_1<a_{j+1}K^{j+1}\le bK^{j+1}.
\end{equation}

We consider the following two cases.

\textbf{Suppose first that $\mathbf{X}$ has rank one over the reals.}
By \Cref{lem:svp-ext-rank-one}, we have $\mathbf{X}=\vec{u}\vec{v}^{\mathsf{T}}$ for nonzero vectors $\vec{u}\in L$ and $\vec{v}\in L^{\otimes j}$.
The induction hypothesis then gives
\[
 \|\mathbf{X}\|_1=\|\vec{u}\|_1\|\vec{v}\|_1
 \ge 2K\cdot a_jK^j\ge a_{j+1}K^{j+1},
\]
contrary to \Cref{eq:svp-ext-short-tensor}.

\textbf{Suppose next that $\mathbf{X}$ has rank at least two over the reals.}
We first find two short columns of $\mathbf{X}$ that are linearly independent modulo $q$.
By its rank, some $2\times2$ minor of $\mathbf{X}$ has a nonzero integer determinant $\Delta$.
Every entry of $\mathbf{X}$ has absolute value at most $\|\mathbf{X}\|_1$, so \Cref{eq:svp-ext-short-tensor} and $j+1\le T$ give
\[
 0<|\Delta|\le2\|\mathbf{X}\|_1^2<2(bK^{j+1})^2\le2(bK^T)^2<q.
\]
Hence $\Delta\not\equiv0\pmod q$, and the two rows of $\mathbf{X}$ through this minor are linearly independent modulo $q$.
Call a column of $\mathbf{X}$ \emph{long} if its $\ell_1$ norm exceeds $bK$.
By \Cref{eq:svp-ext-short-tensor}, fewer than $K^j$ columns are long.
Note that the two rows remain linearly independent modulo $q$ after the long columns are deleted.
Otherwise, some $\F_q$-linear combination of the two rows modulo $q$ is nonzero but vanishes on every column that is not long, so it has fewer than $K^j$ nonzero entries.
This contradicts with the minimum distance of $L^{\otimes j}\bmod q=\mathcal{C}^{\otimes j}$, which is $\lambda(\mathcal{C}^{\otimes j})\ge K^j$.
Thus the columns that are not long span a space of dimension at least two modulo $q$.
We obtain two columns $\vec{u},\vec{v}\in L$ of $\mathbf{X}$ that are linearly independent modulo $q$ and satisfy $\|\vec{u}\|_1,\|\vec{v}\|_1\le bK$.

We next show that the common support of $\vec{u}$ and $\vec{v}$ is large.
Let $J=\{i\in[N]:(u_i,v_i)\ne(0,0)\}$.
Since $\vec{u}\bmod q$ is a nonzero codeword of $\mathcal{C}$ supported in $J$, we have $|J|\ge K$.
Let $t$ be the number of indices $i\in J$ with $|u_i|>4b$ or $|v_i|>4b$.
Each such index contributes at least $4b+1$ to $\|\vec{u}\|_1+\|\vec{v}\|_1\le2bK$, so $t<K/2\le|J|/2$.
At the remaining indices of $J$, the pair $(u_i,v_i)$ is a nonzero integer pair in $[-4b,4b]^2$.
There are fewer than $(8b+1)^2$ such pairs, by an average principle, some pair $(c_1,c_2)\ne(0,0)$ occurs at no fewer than $|J|/(2(8b+1)^2)=\eta|J|$ indices.
Consider the codeword $(c_2\vec{u}-c_1\vec{v})\bmod q$ of $\mathcal{C}$.
Since $|c_1|,|c_2|\le4b<q$, the pair $(c_1,c_2)$ is nonzero modulo $q$, $(c_2\vec{u}-c_1\vec{v})\bmod q$ is nonzero codeword of $\mathcal{C}$ by the linear independence of $\vec{u}$ and $\vec{v}$ modulo $q$.
It remains zero outside $J$ and at every index $i$ with $(u_i,v_i)=(c_1,c_2)$.
Hence
\[
 K\le\|(c_2\vec{u}-c_1\vec{v})\bmod q\|_0\le(1-\eta)|J|,
 \qquad\text{so}\quad |J|\ge\frac{K}{1-\eta}\ge(1+\eta)K.
\]

For each $i\in J$, (the transpose of) row $i$ of $\mathbf{X}$ is a nonzero vector of $L^{\otimes j}$, 
Its transpose is a nonzero vector of $L^{\otimes j}$, 
so it has $\ell_1$ norm at least $a_jK^j$ by the induction hypothesis.
Summing over $i\in J$ gives
\[
 \|\mathbf{X}\|_1\ge|J|\,a_jK^j\ge(1+\eta)a_jK^{j+1}\ge a_{j+1}K^{j+1},
\]
contrary to \Cref{eq:svp-ext-short-tensor}.
This completes the induction.
\end{proof}

We now give the reduction.
It uses the lattice of \Cref{thm:appendix-rs-locally-dense-lattice} in the block matrix of \Cref{thm:CVP-to-SVP}.
In addition, we repeat the last block of rows $K+1$ times.
A lattice vector that violates the MLD constraints modulo $q$ then has at least $K+1$ nonzero coordinates modulo $q$, which gives the bound on the minimum distance modulo $q$ required in \Cref{lem:svp-ext-amplification}.

\begin{theorem}\label{thm:svp-ext-reduction}
Fix a computable real $p\ge1$ and a constant $\gamma\ge1$.
There are an integer $c\ge2$, a computable function $k\mapsto q(k)$ whose values are prime, and a deterministic $\FPT$ many-one reduction from $(2c)$-$\GapMLD_{q(k)}$ to $\gamma$-$\GapSVP_p$.
The output distance parameter depends only on the input parameter $k$.
\end{theorem}

\begin{proof}
We first fix the constants.
Choose an integer $b\ge4$ with $b>(3\gamma)^p$, and let $T$ be the even integer given by \Cref{lem:svp-ext-amplification}.
Since $T$ is fixed and $(1+2/c)^T\to1$ as $c\to\infty$, we can choose an integer $c\ge2$ such that
\begin{equation}\label{eq:svp-ext-gap}
 (3\gamma)^p(1+2/c)^T<b.
\end{equation}
Both choices can be made using a fixed integer upper bound on $(3\gamma)^p$.
Given $k$, put $K=ck$, and let $q(k)$ be the prime returned by the algorithm in \Cref{lem:prime-selection} on input $2(bK^T)^2$.
Thus $q(k)>2(bK^T)^2>2K$.

We next construct the output instance.
Let $(\mathbf{M},\vec{t},k)$ be the input, write $q=q(k)$, and let $\mathbf{M}\in\F_q^{r\times m}$ and $\vec{t}\in\F_q^r$.
We may assume $m\ge1$, since adding a zero column to $\mathbf{M}$ preserves both cases of the promise.
Apply the algorithm in \Cref{thm:appendix-rs-locally-dense-lattice} to $(m,k,K,q)$, that is, with $h=k$ and with $K$ in place of $k$.
This is possible because $K\ge2$ and $q>2K$.
Let $\mathbf{H}$, $\vec{s}$, and $S$ be its outputs, let $N$ be the number of columns of $\mathbf{H}$, and put $\Lambda=\{\vec{v}\in\Z^N:\mathbf{H}(\vec{v}\bmod q)=\vec{0}\}$.
By conditions~\ref{item:app-distance} and~\ref{item:app-center} of that theorem,
\[
 \lambda_1^{(1)}(\Lambda)\ge2K,\qquad
 \lambda(\Lambda\bmod q)\ge K+1,\qquad
 \vec{s}|_S=\vec{0}.
\]
Use \Cref{lem:congruence-kernel-basis} to compute an integer basis $\mathbf{A}\in\Z^{N\times N}$ of $\Lambda$, and let $\mathbf{A}_S$ consist of the rows of $\mathbf{A}$ indexed by $S$, in the specified order.
We identify $\mathbf{M}$ and $\vec{t}$ with their integer lifts with entries in $\{0,\ldots,q-1\}$.
Set $\beta=2K$, let $\mathbf{E}\in\Z^{(K+1)r\times r}$ consist of $K+1$ copies of $\mathbf{I}_r$ stacked vertically, and form
\[
 \mathbf{B}'=
 \begin{pmatrix}
  \mathbf{A} & -\vec{s} & \mathbf{0}\\
  \mathbf{0} & 1 & \mathbf{0}\\
  \beta\mathbf{E}\mathbf{M}\mathbf{A}_S & -\beta\mathbf{E}\vec{t} & \beta q\mathbf{E}
 \end{pmatrix}.
\]
The columns of $\mathbf{B}'$ are linearly independent.
Indeed, in a vanishing linear combination, the second block forces the coefficient of the middle column to be zero.
The first block then forces the coefficients of the first $N$ columns to be zero, since $\mathbf{A}$ is invertible.
The remaining coefficients are zero because $\mathbf{E}$ has linearly independent columns.
Hence $(\mathbf{B}')^{\otimes T}$ is a basis of $\lat(\mathbf{B}')^{\otimes T}$.

To choose the distance parameter, put $R=(K+2k)^{T/p}$.
Since $p$ is computable, we can compute a rational number $a$ with $|a-R|<1/4$, and we set $d'=\lceil a+1/4\rceil$.
Since $R\ge1$, this gives
\begin{equation}\label{eq:svp-ext-threshold}
 R\le d'<R+3/2<3R.
\end{equation}
The reduction outputs $((\mathbf{B}')^{\otimes T},d')$.

We now prove completeness.
Suppose that the input is a YES instance, and take $\vec{x}\in\{0,1\}^m$ with $\|\vec{x}\|_0\le k$ and $\mathbf{M}\vec{x}=\vec{t}$ over $\F_q$.
By condition~\ref{item:app-density} of \Cref{thm:appendix-rs-locally-dense-lattice}, there exists $\vec{y}\in\Z^N$ with $(\mathbf{A}\vec{y}-\vec{s})|_S=\vec{x}$ and
\[
 \|\mathbf{A}\vec{y}-\vec{s}\|_p^p=K-1+2\|\vec{x}\|_p^p=K-1+2\|\vec{x}\|_0,
\]
where the last equality uses $\vec{x}\in\{0,1\}^m$.
Since $\vec{s}|_S=\vec{0}$, we have $\mathbf{A}_S\vec{y}=\vec{x}$ over the integers.
Since $\mathbf{M}\vec{x}\equiv\vec{t}\pmod q$, the vector $\vec{z}=(\vec{t}-\mathbf{M}\vec{x})/q$ is integral, and
\[
 \vec{v} := \mathbf{B}'\begin{pmatrix}\vec{y}\\1\\\vec{z}\end{pmatrix}
 =\begin{pmatrix}\mathbf{A}\vec{y}-\vec{s}\\1\\\vec{0}\end{pmatrix}.
\]
The vector $\vec{v}$ is nonzero, and $\|\vec{v}\|_p^p=1+\|\mathbf{A}\vec{y}-\vec{s}\|_p^p=K+2\|\vec{x}\|_0\le K+2k$.
Hence $\vec{v}^{\otimes T}$ is a nonzero vector of the output lattice with
\[
 \|\vec{v}^{\otimes T}\|_p^p
 =(\|\vec{v}\|_p^p)^T
 \le(K+2k)^T=R^p\le(d')^p.
\]
Thus the output is a YES instance.

We next prove soundness.
Suppose that the input is a NO instance.
We verify the two hypotheses of \Cref{lem:svp-ext-amplification} for the lattice $\lat(\mathbf{B}')$.
Every vector of $\lat(\mathbf{B}')$ has the form
\begin{equation}\label{eq:svp-ext-block-vector}
 \begin{pmatrix}
  \mathbf{A}\vec{y}-\ell\vec{s}\\
  \ell\\
  \beta\mathbf{E}(\mathbf{M}\mathbf{A}_S\vec{y}-\ell\vec{t}+q\vec{z})
 \end{pmatrix},
 \qquad \vec{y}\in\Z^N,\quad\ell\in\Z,\quad\vec{z}\in\Z^r.
\end{equation}
The NO condition controls the first block in the following way.
If $\mathbf{M}\mathbf{A}_S\vec{y}=\ell\vec{t}$ over $\F_q$ and $\ell\not\equiv0\pmod q$, then $\ell^{-1}(\mathbf{A}_S\vec{y}\bmod q)$ is a solution of $\mathbf{M}\vec{x}=\vec{t}$ over $\F_q$.
Since $\vec{s}|_S=\vec{0}$, the NO condition then gives
\begin{equation}\label{eq:svp-ext-block-support}
 \|(\mathbf{A}\vec{y}-\ell\vec{s})|_S\bmod q\|_0
 =\|\mathbf{A}_S\vec{y}\bmod q\|_0>2ck=2K.
\end{equation}

We first show that $\lambda_1^{(1)}(\lat(\mathbf{B}'))\ge2K$.
Suppose that the vector in \Cref{eq:svp-ext-block-vector} has $\ell_1$ norm less than $2K$.
Every entry of its last block is a multiple of $\beta=2K$, so that block is zero.
Since $\mathbf{E}$ has linearly independent columns, we get $\mathbf{M}\mathbf{A}_S\vec{y}-\ell\vec{t}+q\vec{z}=\vec{0}$, and in particular $\mathbf{M}\mathbf{A}_S\vec{y}=\ell\vec{t}$ over $\F_q$.
If $q\nmid\ell$, then \Cref{eq:svp-ext-block-support} gives more than $2K$ nonzero entries in the first block, which is impossible.
Thus $q\mid\ell$, and since $|\ell|<2K<q$, we get $\ell=0$.
The first block is then a vector of $\Lambda$ of $\ell_1$ norm less than $2K$, so it is zero as well.
Hence the vector is zero.

We next show that $\lambda(\lat(\mathbf{B}')\bmod q)\ge K+1$.
Suppose that the vector in \Cref{eq:svp-ext-block-vector} is nonzero modulo $q$.
If $\mathbf{M}\mathbf{A}_S\vec{y}\ne\ell\vec{t}$ over $\F_q$, then the last block has at least $K+1$ nonzero coordinates modulo $q$, because $\beta$ is nonzero modulo $q$ and $\mathbf{E}$ repeats every coordinate $K+1$ times.
If $\mathbf{M}\mathbf{A}_S\vec{y}=\ell\vec{t}$ over $\F_q$ and $q\nmid\ell$, then the first block has more than $2K$ nonzero coordinates modulo $q$ by \Cref{eq:svp-ext-block-support}.
In the remaining case, the second and last blocks vanish modulo $q$.
The first block therefore reduces to a nonzero codeword of $\Lambda\bmod q$, which has Hamming weight at least $K+1$.

Since $q>2(bK^T)^2$, \Cref{lem:svp-ext-amplification} applies to $\lat(\mathbf{B}')$.
Together with $K(1+2/c)=K+2k$, \Cref{eq:svp-ext-gap}, and \Cref{eq:svp-ext-threshold}, it gives
\[
 \bigl(\lambda_1^{(p)}(\lat(\mathbf{B}')^{\otimes T})\bigr)^p
 \ge bK^T
 >(3\gamma)^p(K+2k)^T
 =(3\gamma R)^p
 >(\gamma d')^p.
\]
Thus the output is a NO instance.

We finally bound the running time.
The constants $b$, $T$, and $c$ depend only on $p$ and $\gamma$, and $K$, $q$, and $d'$ are computable from $k$ alone.
By \Cref{thm:appendix-rs-locally-dense-lattice}, computing $\mathbf{H}$, $\vec{s}$, and $S$ takes time $f(q+k+K)(m+1)^{O(1)}$ for a computable function $f$.
By \Cref{lem:congruence-kernel-basis}, computing $\mathbf{A}$ takes time polynomial in the dimensions of $\mathbf{H}$ and $\log q$.
The matrix $\mathbf{B}'$ has $N+1+(K+1)r$ rows and $N+1+r$ columns, and its entries are obtained by integer matrix multiplication.
Since $T$ is a constant, writing down $(\mathbf{B}')^{\otimes T}$ takes time polynomial in the encoding length of $\mathbf{B}'$.
Hence the reduction is deterministic and runs in $\FPT$ time with parameter $k$.
\end{proof}

Thus, we obtain the desired result.

\begin{proof}[Proof of \Cref{thm:svp-ext-hardness}]
Fix $p$ and $\gamma$, and take $c$ and $k\mapsto q(k)$ from \Cref{thm:svp-ext-reduction}.
By \Cref{thm:svp-ext-mld-hard}, $(2c)$-$\GapMLD_{q(k)}$ is $\W[1]$-hard under deterministic $\FPT$ many-one reductions.
Composing these reductions with the reduction in \Cref{thm:svp-ext-reduction} proves the theorem.
\end{proof}

\end{document}